# Materials discovery of stable and non-toxic halide perovskite materials for high-efficiency solar cells

**Authors:** Ryan Jacobs[1], Guangfu Luo[1,2], Dane Morgan[1,*]

[1] Department of Materials Science and Engineering, University of Wisconsin-Madison, Madison, WI 53706, USA

[2] Department of Materials Science and Engineering, Southern University of Science and Technology, Shenzhen 518055, P. R. China

*Corresponding author e-mail: ddmorgan@wisc.edu

## Abstract

Two critical limitations of organic-inorganic lead halide perovskite materials for solar cells are their poor stability in humid environments and inclusion of toxic lead. In this study, high-throughput density functional theory (DFT) methods are used to computationally model and screen 1845 halide perovskites in search of new materials without these limitations that are promising for solar cell applications. This study focuses on finding materials that are comprised of nontoxic elements, stable in a humid operating environment, and have an optimal bandgap for one of single junction, tandem Si-perovskite, or quantum dot-based solar cells. Single junction materials are also screened on predicted single junction photovoltaic (PV) efficiencies exceeding 22.7%, which is the current highest reported PV efficiency for halide perovskites. Generally, these methods qualitatively reproduce the properties of known promising nontoxic halide perovskites that have either been experimentally evaluated or predicted from theory. From a set of 1845 materials, 15 materials pass all screening criteria for single junction cell applications, 13 of which have not been previously investigated, such as $(CH_3NH_3)_{0.75}Cs_{0.25}SnI_3$, $((NH_2)_2CH)Ag_{0.5}Sb_{0.5}Br_3$, $CsMn_{0.875}Fe_{0.125}I_3$, $((CH_3)_2NH_2)Ag_{0.5}Bi_{0.5}I_3$, and $((NH_2)_2CH)_{0.5}Rb_{0.5}SnI_3$. These materials, together with others predicted in this study, may be promising candidate materials for stable, highly efficient, and non-toxic perovskite-based solar cells.

**Table of Contents Figure**

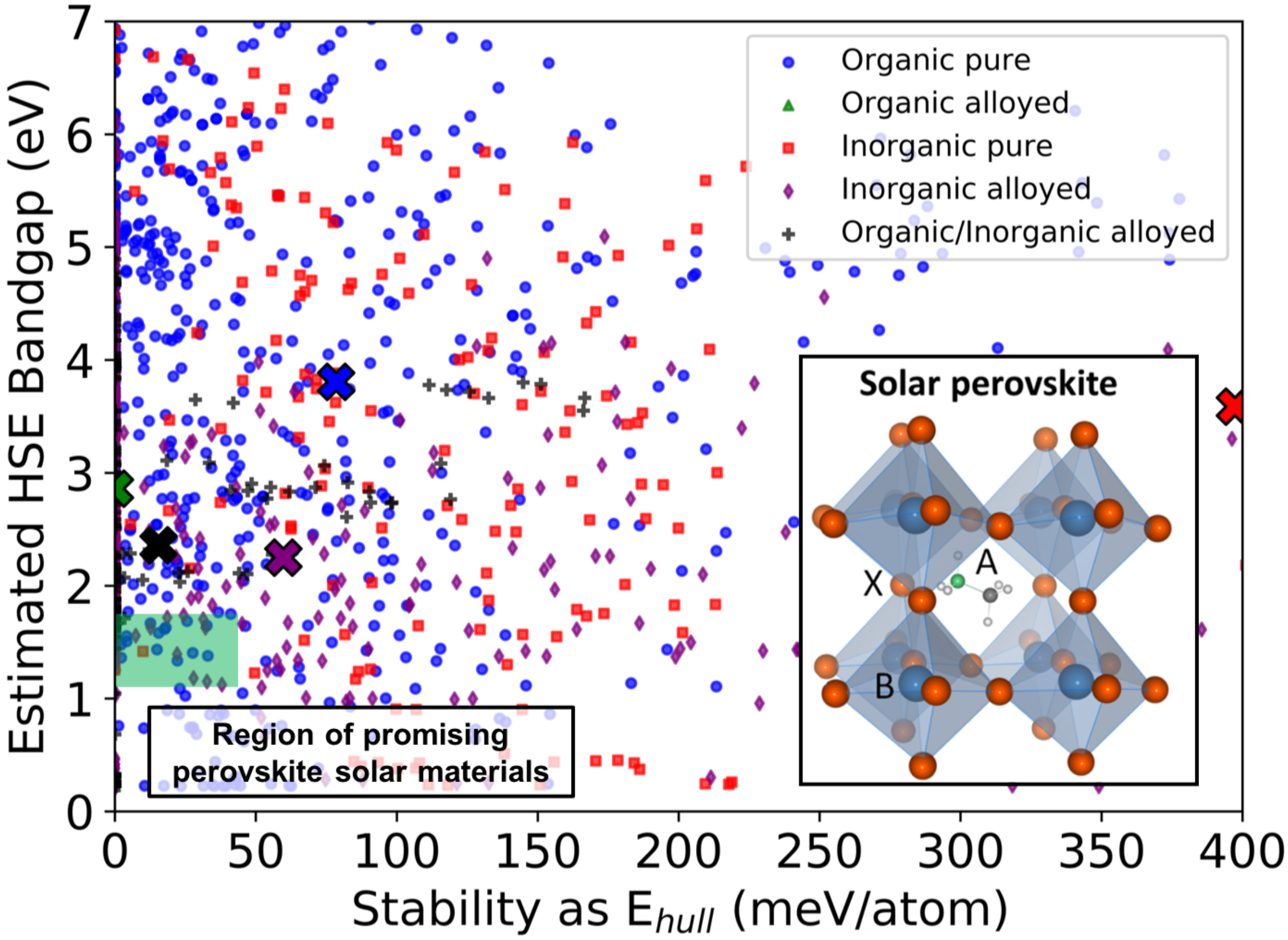

# 1. Introduction

The first report of a hybrid organic-inorganic halide perovskite (general formula $ABX_3$, $A$ = organic 1+ molecule, $B$ = 2+ metal cation, and $X$ = 1- halogen anion) was made in 1978 by Weber, when both the $CH_3NH_3SnBr_xI_{3-x}$ ($x$ = 0-3) and $CH_3NH_3PbX_3$ ($X$ = Br, Cl, I) perovskite systems were successfully synthesized.[1,2] In 2009, the potential of halide perovskites for use in photovoltaic (PV) solar cells was realized, when a solar cell made from $CH_3NH_3PbI_3$ was found to have an energy efficiency of 3.8%.[3] Since then, research in perovskite solar cells has exploded, and extremely rapid progress has been made in enhancing the efficiency of these materials, with the record-holder residing at 22.7% in 2017.[4] Many properties related to the excellent performance of this class of materials have been uncovered and investigated, including tunable direct bandgap,[5–7] excellent light absorption,[3,7,8] very long electron and hole diffusion lengths greater than 1 μm,[9–11] and low quantity of defects acting as recombination centers.[12–15] Halide perovskites are particularly promising for commercializable solar cells as a replacement for, or in complement to, silicon-based cells because they can display high PV efficiencies, are comprised of cheap and abundant precursor materials, and are amenable to inexpensive, scalable synthesis procedures such as solution-based methods.[16,17] Many recent investigations have sought to go beyond employing halide perovskites solely as single junction PV materials. A promising avenue to realize efficiencies not limited by the single junction Shockley–Queisser limit of 34% is through the production of Si-perovskite tandem solar cells. Perovskite tandem cell efficiencies have reached 25%, higher than the record-holding single junction Si and perovskite materials.[18–24] In addition, the inorganic halide perovskites have received significant attention in the form of colloidal quantum dot films, which have application both for light emitting diodes and solar cells.[25–29] Despite the excellent reported efficiencies and other attractive attributes like low manufacturing cost, there are two main issues with the highest performing and most widely studied halide perovskites, $CH_3NH_3PbI_3$ ($CH_3NH_3^+$ is methylammonium, henceforth MA) and $(NH_2)_2CHPbI_3$ ($(NH_2)_2CH^+$ is formamidinium, henceforth FA). These issues are that they contain toxic Pb and their performance deteriorates quickly over time in humid environments, which could limit their successful application in commercial solar cells.[16,30–32]

Previous studies have sought to address the toxicity problem by substituting Pb with Sn or Ge in $MAPbI_3$ and $FAPbI_3$.[16,30,33,34] Doping with Sn or Ge both removes the toxicity issue and increases the stability of the perovskite in humid air, but also substantially lowers its efficiency.

While Sn-containing compounds show better stability with regard to decomposition in humid air, these compounds are still unstable and the reduced efficiency is likely related to the Sn oxidation from $Sn^{2+}$ to $Sn^{4+}$.[32] The Ge-based perovskites also have stability issues due to oxidation of Ge from $Ge^{2+}$ to $Ge^{4+}$,[32,33] which can explain the lower efficiency of Ge-containing perovskites.

In addition to addressing the toxicity issue, there have been many reports of methods to enhance the stability of perovskites based on $MAPbI_3$ and $FAPbI_3$. In addition to doping Sn or Ge on the *B*-site as discussed above, doping Cs and Rb in place of MA and/or FA on the *A*-site, doping of Br and Cl in place of I on the *X*-site, and simultaneous doping of Sn and Ge on the *B*-site have improved the stability in humid environments.[6–8,30,35–41] In addition to compositional refinement, there have been manufacturing-level device engineering approaches such as device encapsulation, film morphology tuning, and modification of the charge collection materials to protect the solar cells from reacting with humid air and enhance their photocurrent stability over long periods of time.[32,42–44] While device encapsulation appears to be one method to alleviate the issue of material stability, such encapsulation may increase the manufacturing cost and overall device size and weight. The latter two items are of particular concern for the use of perovskite solar cells in non-stationary applications.

There have been multiple computational studies aimed at discovering new halide perovskites and double perovskites which are stable and have an optimal bandgap. Here, we summarize the main findings of the most recent and relevant works. First, the work of Krishnamoorthy *et al.*[33] examined the purely inorganic composition space of 360 materials and suggested that $CsGeI_3$ (and their purely MA- and FA-based analogues) are promising materials based on their calculated bandgap and thermodynamic stability. Experimental data from the same work showed that test devices containing $CsGeI_3$ and $MAGeI_3$ exhibited short circuit currents on par with Sn-containing halide perovskites, but lower open circuit voltages. Thermogravimetric analysis data showed that the Ge-containing perovskites were thermodynamically stable under device working conditions, however the formation of $Ge^{4+}$ was to blame for the poor open circuit voltage. The formation process of $Ge^{4+}$ remains an open research problem in these materials. Second, the work of Filip and Giustino[45] investigated a pool of 116 $CsBX_3$ inorganic halide perovskites with homovalent *B*-site substitution of toxic Pb. They found that Mg could be a promising dopant to at least partially substitute for Pb in these materials to reduce the toxicity and tune the bandgap. Third, the work of Nakajima and Sawada[46] examined a large composition space of 11,025 total perovskite and double perovskite materials. However, most of these 11,025 compounds were found to be metallic, and just 2143 materials were found to be semiconducting and thus relevant for solar cell applications. Their composition space focused on *A*-site species containing Cs, MA and FA, *X*-site species of Cl, Br and I and 49 different *B*-site elements. Of this large composition space, they found 12 promising perovskite compounds based on their calculated Density Functional Theory (DFT)-HSE level bandgap values. This set of promising compounds contains previously known promising materials like $CsSnI_3$, $CsSnBr_3$, $CsGeBr_3$, plus new materials such as $MASiI_3$ and $FAAuI_3$. Finally, a study aimed at creating a large database of halide perovskites by Kim *et al.*[47] examined 1346 materials with 16 different organic cations on the *A*-site, *B*-site elements of Ge, Sn and Pb and the *X*-site species F, Cl, Br and I. This study by Kim *et al.* is, to our knowledge, the only study besides the present one which has performed high-throughput DFT calculations on perovskites containing a variety of organic molecules on the *A*-site.

In addition to screening traditional perovskite structures for suitable PV materials, there have also been recent studies aimed at finding promising double halide perovskite materials ($A_2BB'X_6$ structure). First, Volonakis *et al.*[48] explored heterovalent *B*-site doping by modeling nine double halide perovskite ($Cs_2BB'X_6$) materials where the *B*-site was doped with equal concentrations of either $B = Bi^{3+}$ or $Sb^{3+}$ together with one of $B' = Cu^{1+}$, $Ag^{1+}$ or $Au^{1+}$. They found that this class of materials possesses a wide range of bandgaps (from 0 to 2.7 eV using hybrid PBE0 calculations) and have generally low carrier effective masses. While no stability calculations were conducted, they succeeded in synthesizing $Cs_2BiAgCl_6$ using bulk solid-state reaction methods. Second, a study by Cai *et al.*[49] examined trends in the bandgap and stability of a set of 81 double perovskite compositions with *A* = K, Rb, and Cs, *B* = Si, Ge, Sn, Pb, Ni, Pd, Pt, Se, and Te and *X* = Cl, Br and I. Their stability analysis showed that numerous materials with *B* = Pd, Pt, Se, and Te are expected to be stable (or nearly so) for all species considered. Many of these double perovskites have been experimentally synthesized, and $Cs_2PtI_6$ stands out as a promising compound due to it being stable and possessing an HSE-calculated bandgap of 1.47 eV (1.34 eV with spin-orbit coupling effects included), though it is worth noting that the inclusion of Pt would likely not be feasible for large scale commercialization due to prohibitive costs. Third, Zhao *et al.*[50] identified 11 optimal double halide perovskites from a pool of 64 materials examined, with candidates such as $Cs_2BiAuBr_6$ and $Cs_2SbInCl_6$ possessing optimal stability, bandgap, electron and hole effective masses and low exciton binding energy. Finally, the work of Nakajima and Sawada[46] discussed above also examined double perovskite materials. Of the large set of 2243 semiconducting materials they examined, they found 40 promising double perovskite materials based on their DFT-HSE level bandgap values. By examining the compositions of these promising double perovskite materials, *B*-site elements of Si, Au and In appear frequently; the inclusion of Au and In in the double perovskite structure is in agreement with the study discussed above by Zhao *et al.*[50]

Although a number of computational screening studies have been conducted, the present work is novel for three reasons. First, the screening analysis employed here is more complete as it (i) incorporates stability assessment by using convex hull analysis under a realistic solar cell operating environment, which is more accurate for determining stability than approaches employed in previous studies, and (ii) incorporates PV efficiency calculations to place a more stringent screening criterion beyond simply analyzing bandgaps. Second, we have considered a more diverse composition space than previous studies, consisting of both organic molecules and inorganic elements, and also a variety of alloying schemes for the perovskite *A*-, *B*- and *X*-site. For example, we have considered numerous compositions of the *A*-, *B*- and *X*-site, including combined *A*-site/*B*-site and *B*-site/*X*-site alloying and mixtures of inorganic and organic species on the *A*-site. Finally, we have expanded our search beyond the usual goal of new materials for single junction solar cells, and provide recommendations of promising materials for Si-perovskite tandem cells and quantum dot applications.

In this work, we have used high-throughput DFT methods enabled by the Materials Simulation Toolkit (MAST)[51] to simulate and screen a total of 1845 different halide perovskite compositions in search of non-toxic, stable, and high efficiency solar cell materials. Additional information on the different composition spaces that we explored is provided in **Section S1** of the **Supplementary Information (SI)** and a complete list of all materials is provided in the spreadsheet as part of the **SI** and available on Figshare (see link in section summarizing the **Supplementary Information**). To evaluate the potential of each halide perovskite as a solar cell material, we conducted a five-step screening evaluation to down-select promising materials based on a series of criteria. **Figure 1** presents an overview of the sequential screening and elimination criteria used to discover new promising compounds. First, we removed compounds that contain toxic elements (**Section 2.1**). Second, we calculated the stability using convex hull analysis to obtain the energy above the convex hull ($E_{hull}$). Convex hull analysis has been successfully used in numerous other materials screening studies to obtain qualitative insight of materials stability and synthesizability, including for halide perovskites.[33,49,52–54] We grouped compounds based on their stability, and since DFT errors and metastability make the optimal criteria for stability uncertain, we used specific compound energies with known issues as guides. Specifically, compounds were categorized as stable ($E_{hull} \leq 15$ meV/atom, where 15 meV/atom is the stability value for $MAPbI_3$), metastable (15 meV/atom < $E_{hull} \leq 46$ meV/atom, where 46 meV/atom is the stability value for $MASnI_3$), or unstable, and we removed unstable compounds (**Section 2.2**). Additional details on performing the convex hull analysis and the stability screening are provided in **Section 4** and **Section S4** of the **SI**. Third, we calculated the PBE-level bandgap and used this value to estimate the HSE-level bandgap, and removed materials with predicted HSE bandgaps outside the range of 1.1-1.7 eV (**Section 2.3**). Fourth, we explicitly calculated HSE-level bandgaps and removed materials with HSE bandgaps outside the range of 1.1-1.7 eV, 1.9-2.3 eV, and 0.8-1.4 eV for applications in single junction, Si-perovskite tandem cell, and quantum dot-based applications, respectively (**Section 2.4** for single junction**, Section 2.6** for tandem and quantum dot). Finally, for just the materials which passed the single junction bandgap requirement, we calculated the PV efficiency assuming a 0.5 µm film thickness,[5,8,9]and removed materials with PV efficiencies below 22.7%, which is the current best-reported halide perovskite cell efficiency[4] (**Section 2.5**). Additional details about how each elimination criterion was selected are provided in **Section S2** of the **SI**. After applying the screening criteria outlined in **Figure 1** on our initial pool of 1845 halide perovskites, there are 15 promising compounds that pass all screening criteria for new single junction solar perovskite materials, 13 of which have not, to our knowledge, been previously experimentally or theoretically explored. Based on the calculated bandgaps of the 62 materials that pass screening criteria 1, 2 and 3 (material sets 4A and 4B), 13 and 26 of them are also promising for Si-perovskite tandem and quantum dot applications, respectively.

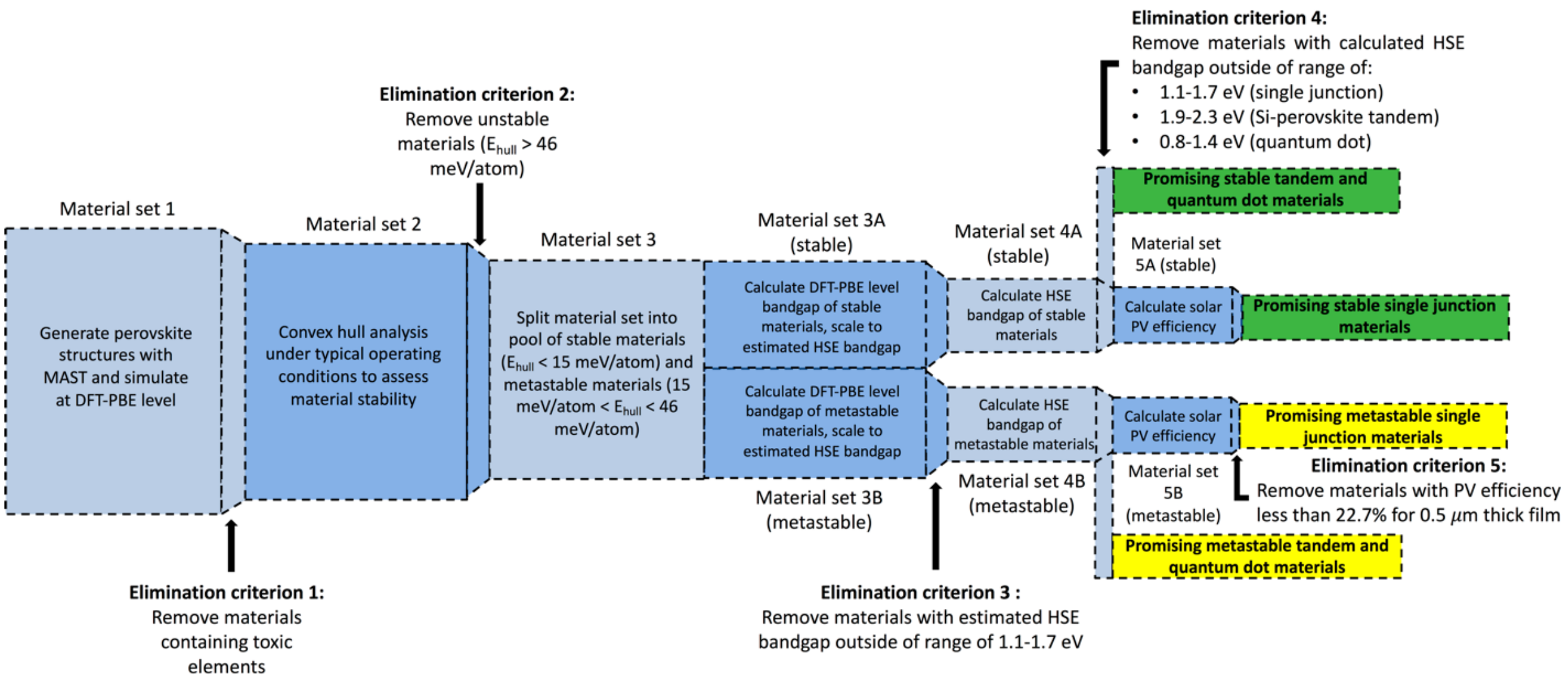

**Figure 1**. Summary of screening and elimination criteria used in this study. For materials sets 1-5, the listed elimination criteria were invoked to reduce the number of candidate perovskite materials. These promising materials are comprised of nontoxic elements, have stability values that are either classified as stable or metastable, possess estimated and calculated HSE bandgaps within the optimal bandgap range of 1.1-1.7 eV, 1.9-2.3 eV, and 0.8-1.4 eV for single junction cells, Si-perovskite tandem cells, and quantum dot applications, respectively. For the case of promising single junction cell materials, our promising compounds also possess calculated ideal PV efficiencies of at least 22.7% for a 0.5 μm thick film.

# 2. Results and Discussion

## 2.1. Toxic element screening

Material set 1 is comprised of 1845 distinct halide perovskite compositions. Of this initial pool of 1845 compounds, 39 compounds did not adequately converge in our DFT calculations. The inability to obtain a converged perovskite structure may suggest these compositions are not stable in the perovskite phase. These non-converged calculations were removed from material set 1, thus resulting in 1806 remaining materials. As shown in **Figure 1**, all materials were simulated at the PBE level, and then our first elimination criterion of removing compounds that contain toxic Pb, Cd and Be was applied. A central point of this study is the discovery of perovskite materials comprised of nontoxic elements, yet we have simulated compounds that contain the toxic elements Pb, Cd and Be on the *B*-site. Our rationale for simulating materials containing these toxic elements is in case future advances in material recycling and solar module manufacturing minimize the Pb toxicity issue to the point where it is acceptable to include toxic elements in perovskite-based solar cells. In all, there are 353 materials that contain toxic elements, therefore removing these compounds from consideration results in material set 2 containing 1453 compounds.

### 2.2. Thermodynamic stability screening

The second step of our screening criteria is to assess the thermodynamic stability of all compounds in material set 2 (see **Figure 1**). To do this, we used the phase stability tools contained in the Pymatgen software package[55] to conduct convex hull analysis under typical solar cell operating conditions of $T = 298$ K, $p(O_2) = 0.2$ atm and a relative humidity of 30% to calculate the energy above the convex hull $E_{hull}$ for each system.[55] **Figure 2** is a histogram of stability values for all materials comprising material set 2. Based on the elimination criterion for material set 2 in **Figure 1**, the materials in **Figure 2** are binned according to whether they are predicted to be stable (green bars, $E_{hull} \leq 15$ meV/atom above convex hull, where this limit is set by the stability value of $MAPbI_3$), metastable (yellow bars, $15 < E_{hull} \leq 46$ meV/atom above convex hull, where the upper limit is set by the stability value of $MASnI_3$), or unstable (red bars, $E_{hull} > 46$ meV/atom above convex hull). From this analysis of 1453 compounds, 720 are predicted to be stable, 172 are metastable, and 561 are unstable and are eliminated from further consideration. Thus, material set 3A (3B) consists of a total of 720 (172) stable (metastable) materials, for a total of 892 materials to conduct bandgap screening on.

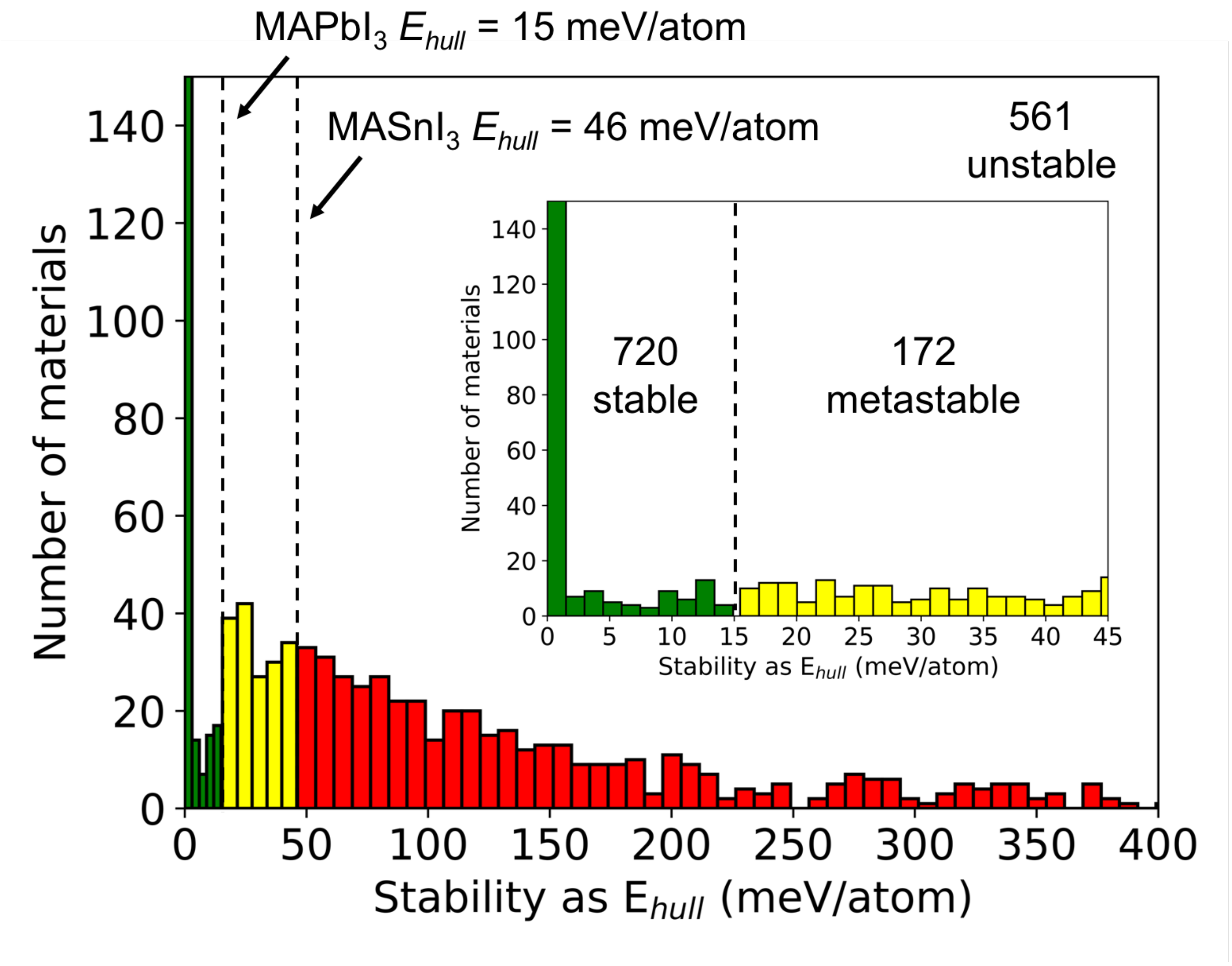

**Figure 2.** Histogram cataloging the stability of all materials contained in material set 2. The green, yellow, and red colored bars indicate materials that pass the stability criteria of $E_{hull}$ ≤ 15 meV/atom, metastability criteria of 15 meV/atom < $E_{hull}$ ≤ 46 meV/atom, and unstable criteria of $E_{hull}$ > 46 meV/atom, respectively.

### 2.3. Screening based on estimated HSE bandgap

Our bandgap screening consists of two steps (see **Figure 1**). First, we calculate the PBE-level bandgap for each of the 720 (172) stable (metastable) materials in material set 3A (material set 3B). Next, these PBE-level bandgaps are scaled using the scaling relationship $E_{gap}^{HSE} = (1.2607 \pm 0.0035)E_{gap}^{PBE} + (0.2246 \pm 0.0117)$eV, where $E_{gap}^{HSE}$ and $E_{gap}^{PBE}$ are the HSE and PBE bandgaps, respectively. This scaling relationship was obtained from the best-fit line between calculated HSE and PBE bandgaps from the work of Kim *et al.*[47] and Pilania *et al.*,[56] who collectively computed bandgaps of nearly 1600 hybrid perovskite systems (see **Figure S2** in **Section S2** of the **SI** for a plot of this scaling relationship). The mean error, mean absolute error, and standard deviation between the actual and predicted (from the scaling relationship above) HSE bandgap values are $-6.9\times10^{-6}$ eV, 0.11 eV and 0.16 eV, respectively. The very small mean error demonstrates the distribution of estimated HSE bandgaps is spread evenly above and below the best-fit line. The magnitude of the standard deviation is one source of error for our estimated HSE bandgap elimination criterion. In addition to error resulting from the PBE vs. HSE bandgap fit described above, there is some error between HSE-calculated bandgaps and their experimental values. Previous studies analyzing the accuracy of HSE bandgaps have found that mean absolute errors (standard deviations) between HSE and experimental bandgaps for a diverse set of covalent semiconductors and insulators are about 0.23-0.28 (0.2-0.36, averaging to 0.25 between studies) eV.[57–60] Thus, we obtain an approximate acceptable range of estimated HSE bandgap values by independently adding the error of the PBE bandgap relative to the HSE bandgap (taken as one standard deviation, 0.16 eV) and the error of the HSE bandgap relative to experimental bandgap (taken as 0.25 eV), which results in a range of +/- 0.3 eV. Thus, our estimated HSE bandgap screening criterion value around the ideal 1.4 eV bandgap for a single junction solar cell[61,62] is therefore taken as 1.4 +/- 0.3 eV, i.e., 1.1 eV to 1.7 eV. This choice of estimated HSE bandgap range has the benefit of resulting in a tractable number of HSE bandgap calculations (see **Section S2** of the **SI** for additional details). The procedure used here of first screening materials by estimating their HSE bandgaps with the linear scaling relationship followed by explicit HSE bandgap calculations will tend to minimize the occurrence of false positive results.

**Figure 3** contains histograms of PBE bandgaps for the stable materials (**Figure 3A**) and metastable materials (**Figure 3B**) after the stability screening. Histograms of the estimated HSE bandgaps for the stable materials (**Figure 3C**) and metastable materials (**Figure 3D**) are also included. It was found that the pool of 720 (172) stable (metastable) compounds of material set 3A (3B) down-select to 51 (11) materials passing the estimated HSE bandgap criterion. Thus, material set 4A (4B) consists of 51 (11) stable (metastable) materials which constitutes the set of materials to be further evaluated with explicit HSE bandgap calculations.

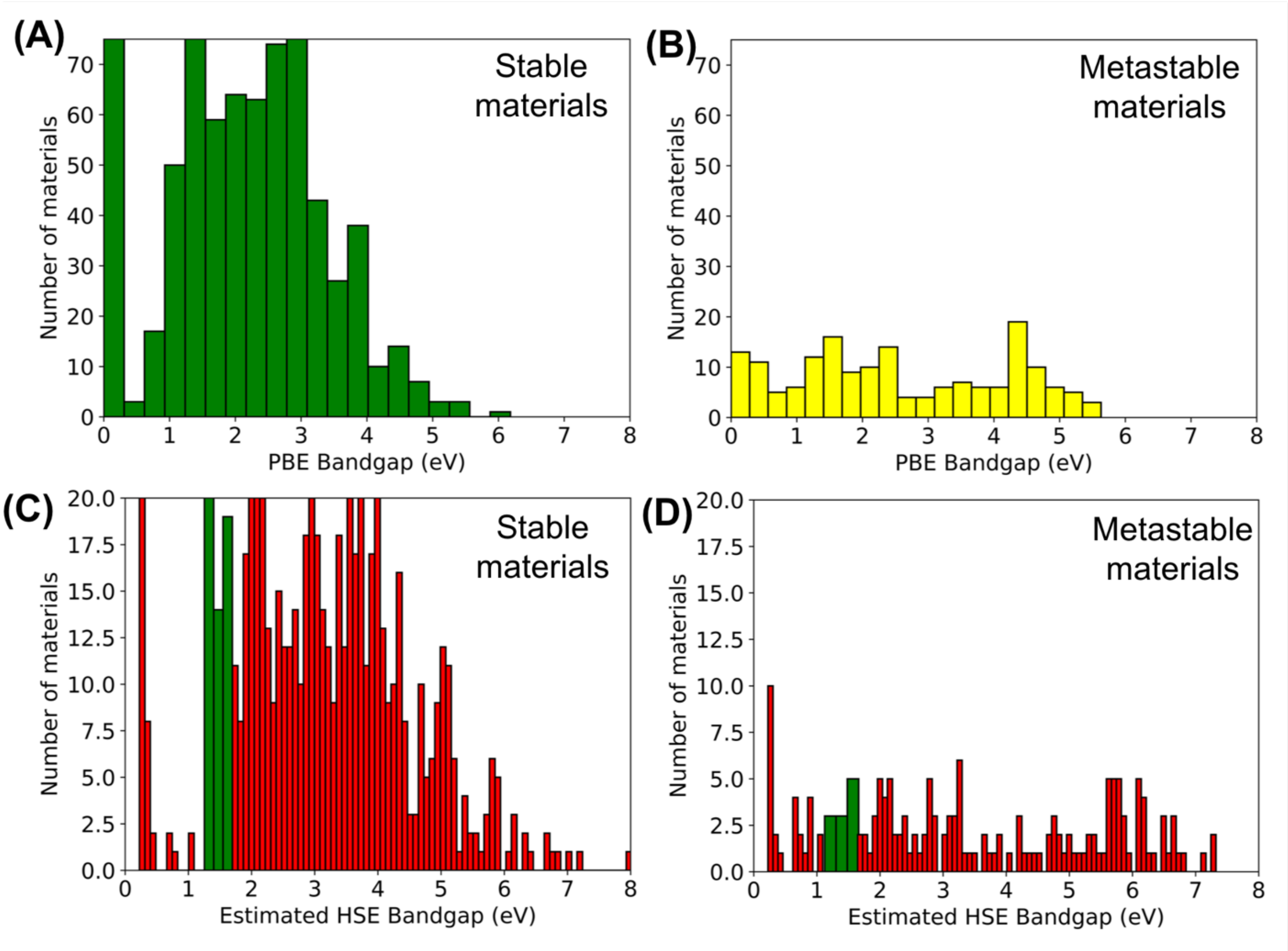

**Figure 3.** Histograms of PBE bandgaps for stable (A) and metastable (B) materials that passed the stability screening step and histograms of estimated HSE bandgaps for stable (C) and metastable (D) materials. In (C) and (D), the green (thicker) bars represent materials with estimated HSE bandgap within the range of 1.1-1.7 eV, while the red (thinner) bars represent materials outside the estimated HSE bandgap range of 1.1-1.7 eV. In all, 51 (11) stable (metastable) materials pass the estimated HSE bandgap criterion.

### 2.4. Screening based on explicit HSE bandgap calculations

After elimination criteria 3 (**Section 2.3)**, we have identified 62 halide perovskite materials (material sets 4A and 4B) which pass the stability and estimated HSE bandgap screening criteria. For these sets of materials, we have explicitly calculated the HSE bandgap to create material set 5A (stable compounds) and 5B (metastable compounds). The stability, calculated PBE bandgap, estimated HSE bandgap, and calculated HSE bandgap for all 62 materials in set 4A and 4B are tabulated in **Table 1**. **Table 1** shows that explicit calculation of the HSE bandgaps for all 51 (11) materials in material set 4A (4B) results in 28 (4) of them passing the calculated HSE bandgap screening criterion, thus forming material set 5A (5B), which will be subject to PV efficiency screening in **Section 2.5**. From comparing the difference between the calculated and estimated HSE bandgaps for the materials in **Table 1**, we find that the mean error and mean absolute error are 0.10 and 0.28 eV, respectively. These errors are larger than the mean error ($-6.9\times10^{-6}$ eV) and mean absolute error (0.11 eV) obtained from the linear fit discussed above (see **Figure S2**). However, these larger errors are to be expected, which are likely the result of the bandgaps in **Table 1** occupying only a very narrow range of the values fit in **Figure S2**, and the inclusion of material compositions in **Table 1** that were not contained in the original fit.

**Table 1.** Summary of 51 (11) stable (metastable) materials in material set 4A and 4B. Of these, 28 (4) stable (metastable) materials pass the calculated HSE bandgap screening criterion, thus forming material set 5A and 5B. The material ID numbers provide a reference to the catalogued materials information in the spreadsheet included as part of the **SI**.

| **Stable materials** | | | | | |
|---|---|---|---|---|---|
| **Material** | **ID #** | **$E_{hull}$ (meV/atom)** | **PBE bandgap (eV)** | **Estimated HSE bandgap (eV)** | **Calculated HSE bandgap (eV)** |
| $CsBi_{0.5}Ag_{0.5}I_3$ | 17 | 12.7 | 1.11 | 1.63 | 1.55 |
| $((CH3)_2NH_2)Bi_{0.5}Ag_{0.5}I_3$ | 161 | 0 | 1.06 | 1.56 | 1.53 |
| $FAAg_{0.5}Sb_{0.5}Br_3$ | 197 | 0 | 1.09 | 1.60 | 1.56 |
| $Cs_{0.25}Na_{0.75}CoI_3$ | 221 | 0 | 1.13 | 1.65 | 2.37 |
| $RbFeI_3$ | 243 | 0 | 1.00 | 1.48 | 1.76 |
| $Cs_{0.25}Rb_{0.75}FeI_3$ | 246 | 0 | 1.05 | 1.55 | 1.73 |
| $Cs_{0.875}Na_{0.125}MnI_3$ | 257 | 0 | 0.92 | 1.38 | 1.35 |
| $Cs_{0.5}Na_{0.5}MnI_3$ | 259 | 7.4 | 1.06 | 1.56 | 1.58 |
| $Cs_{0.75}Na_{0.25}MnI_3$ | 260 | 0 | 0.93 | 1.39 | 1.33 |
| $Cs_{0.875}Rb_{0.125}MnI_3$ | 262 | 0 | 0.93 | 1.39 | 1.34 |
| $RbMnI_3$ | 263 | 0 | 0.93 | 1.40 | 1.34 |
| $Cs_{0.5}Rb_{0.5}MnI_3$ | 264 | 0 | 0.93 | 1.40 | 1.34 |
| $Cs_{0.75}Rb_{0.25}MnI_3$ | 265 | 0 | 0.93 | 1.39 | 1.34 |
| $Cs_{0.25}Rb_{0.75}MnI_3$ | 266 | 0 | 0.93 | 1.40 | 1.34 |
| $CsMn_{0.875}Co_{0.125}I_3$ | 267 | 0 | 0.92 | 1.39 | 1.75 |
| $CsMn_{0.5}Co_{0.5}I_3$ | 269 | 0 | 0.98 | 1.46 | 2.58 |
| $CsMn_{0.75}Co_{0.25}I_3$ | 270 | 0 | 0.93 | 1.40 | 1.50 |
| $CsMn_{0.875}Fe_{0.125}I_3$ | 272 | 0 | 0.91 | 1.38 | 1.33 |
| $RbMn_{0.875}Co_{0.125}I_3$ | 348 | 0 | 0.91 | 1.38 | 1.31 |
| $RbMn_{0.75}Co_{0.25}I_3$ | 351 | 0 | 0.94 | 1.41 | 1.31 |
| $RbMn_{0.875}Fe_{0.125}I_3$ | 353 | 0 | 0.91 | 1.37 | 1.35 |
| $RbMn_{0.5}Fe_{0.5}I_3$ | 355 | 0 | 0.88 | 1.33 | 1.46 |
| $RbMn_{0.75}Fe_{0.25}I_3$ | 356 | 0 | 0.90 | 1.37 | 1.34 |
| $(CH_3CH_2)CoI_3$ | 387 | 0 | 1.06 | 1.56 | 2.27 |
| $MAFeI_3$ | 635 | 0 | 1.05 | 1.54 | 1.89 |
| $MAMnI_3$ | 647 | 0 | 0.82 | 1.25 | 1.27 |
| $((CH_3)_2CH)BaI_3$ | 732 | 2.2 | 0.90 | 1.36 | 1.90 |
| $((CH_3)_2CH)CoI_3$ | 748 | 0 | 1.11 | 1.63 | 2.09 |
| $((CH_3)_2CH)FeI_3$ | 756 | 0 | 1.04 | 1.54 | 1.78 |
| $((CH_3)_2CH)MnI_3$ | 768 | 0 | 0.97 | 1.45 | 1.78 |
| $((CH_3)_2CH)SrI_3$ | 780 | 2.2 | 0.99 | 1.48 | 2.01 |
| $((CH_3)_2CH)ZnI_3$ | 784 | 12.6 | 1.15 | 1.67 | 2.14 |
| $((CH_3)_2NH_2)MnI_3$ | 828 | 0 | 1.04 | 1.54 | 1.78 |
| $((CH_3)_3C)CoI_3$ | 868 | 0 | 0.99 | 1.47 | 2.08 |
| $((CH_3)_3C)FeI_3$ | 876 | 0 | 1.06 | 1.56 | 2.03 |
| $((CH_3)_3C)MnI_3$ | 888 | 0 | 1.04 | 1.54 | 2.07 |
| $((CH_3)_3C)ZnI_3$ | 904 | 4.2 | 1.02 | 1.51 | 2.12 |
| $CsMnI_3$ | 1008 | 0 | 0.92 | 1.38 | 1.33 |
| $KFeI_3$ | 1056 | 0 | 0.86 | 1.31 | 1.64 |
| $KMnI_3$ | 1068 | 10.0 | 0.95 | 1.42 | 1.33 |
| $MA_{0.875}Cs_{0.125}SnI_3$ | 1680 | 0.7 | 1.15 | 1.67 | 1.18 |
| $MA_{0.75}Cs_{0.25}FeI_3$ | 1686 | 0 | 1.12 | 1.64 | 1.77 |
| $MA_{0.75}Cs_{0.25}SnI_3$ | 1695 | 0 | 1.17 | 1.70 | 1.20 |
| $MA_{0.5}Cs_{0.5}MnI_3$ | 1707 | 0 | 0.97 | 1.44 | 1.44 |
| $MA_{0.5}Cs_{0.5}SnI_3$ | 1710 | 0 | 1.12 | 1.64 | 1.08 |
| $MA_{0.875}Rb_{0.125}FeI_3$ | 1716 | 0 | 1.05 | 1.54 | 1.73 |
| $MA_{0.75}Rb_{0.25}FeI_3$ | 1731 | 0 | 1.10 | 1.62 | 1.82 |
| $MA_{0.75}Rb_{0.25}SnI_3$ | 1740 | 0 | 1.17 | 1.69 | 1.21 |
| $MA_{0.5}Rb_{0.5}FeI_3$ | 1755 | 0 | 1.12 | 1.64 | 1.07 |
| $FA_{0.5}Cs_{0.5}SnI_3$ | 1800 | 0 | 1.09 | 1.60 | 1.11 |
| $FA_{0.5}Rb_{0.5}SnI_3$ | 1845 | 0 | 1.12 | 1.64 | 1.12 |
| **Metastable materials** | | | | | |
| **Material** | **ID #** | ***$E_{hull}$* (meV/atom)** | **PBE bandgap (eV)** | **Estimated HSE bandgap (eV)** | **Calculated HSE bandgap (eV)** |
| $CsBi_{0.5}Cu_{0.5}I_3$ | 19 | 38.0 | 0.71 | 1.12 | 1.39 |
| $CsSb_{0.5}Ag_{0.5}I_3$ | 21 | 27.8 | 0.76 | 1.18 | 1.02 |
| $RbCu_{0.5}Bi_{0.5}Cl_3$ | 59 | 45.1 | 1.14 | 1.66 | 2.23 |
| $RbAg_{0.5}Bi_{0.5}I_3$ | 65 | 32.1 | 1.11 | 1.63 | 1.53 |
| $RbSb_{0.5}Ag_{0.5}I_3$ | 69 | 32.6 | 0.74 | 1.15 | 0.97 |
| $RbMn_{0.5}Co_{0.5}I_3$ | 350 | 22.1 | 0.93 | 1.39 | 1.94 |
| $((CH_3)_2CHNH_3)CuF_3$ | 691 | 22.1 | 0.88 | 1.34 | 4.58 |
| $((CH_3)_2CH)CaI_3$ | 740 | 18.0 | 1.12 | 1.63 | 2.11 |
| $((CH_3)_3C)GeI_3$ | 880 | 27.6 | 0.94 | 1.41 | 1.68 |
| $((CH_3)_3C)MgI_3$ | 884 | 17.6 | 1.10 | 1.62 | 2.19 |
| $FASnI_3$ | 1317 | 32.6 | 1.13 | 1.65 | 1.23 |

## 2.5. Screening based on solar PV efficiency and summary of the most promising new compounds

Finally, we conclude our screening by calculating the PV efficiency of the 28 (4) materials from material set 5A (5B) obtained above. **Table 2** summarizes the calculated PV efficiencies for all materials in set 5A and 5B, sorted from high to low calculated PV efficiency. As shown in **Figure 1**, our final elimination criterion was to remove compounds with predicted PV efficiency less than 22.7%. **Table 2** shows that 12 (3) stable (metastable) materials pass this screening criterion, for a final total of 15 promising compounds. To our knowledge, 13 of these 15 materials have not been previously studied and thus constitute the set of new promising compounds worth further investigation. We recognize that while this PV efficiency value of 22.7% is a very stringent condition for determining whether a material may be a promising light absorber, the nature of the PV efficiency calculations conducted in this study are highly idealized, and thus represent an upper bound of the expected efficiency of the actual material in a real device.

Here, we remark on additional promising halide perovskites we found from examining the work of Kim *et al.*,[47] which is separate from the screening process discussed above. In addition to the promising compounds based on HSE bandgap screening reported in **Table 1**, the work of Kim *et al.* also provides a large database of halide perovskite bandgaps. The work of Kim *et al.* doesn't make any specific mention of possible promising compounds from the creation of their database, however from examining their reported DFT-HSE bandgap values, new materials such as $NH_2NH_3SnI_3$ (bandgap of 1.52 eV, $NH_2NH_3$ is hydrazinium), $C_3H_6N_2SnI_3$ (bandgap of 1.58 eV, $C_3H_6N_2$ is azetidinium) and $C(NH_2)_3SnI_3$ (bandgap of 1.64 eV, $C(NH_2)_3$ is guanidinium) may also be worth further examination, though we note that the stability of these compounds was not reported.

The 13 new compounds discussed in this section are promising based on the screening criteria employed in this work, namely the calculated thermodynamic stability, bandgap, and PV efficiency. In addition to these quantities, there are other physical properties that are important to consider for efficient solar cell operation which we have not considered here. These properties include, but are not limited to: the presence of defect levels in the gap,[12,13,63–68] electron and hole effective masses and mobilities,[45,48,49,69] electron-hole pair lifetimes,[10,70–72] and the electronic orbitals comprising the valence and conduction bands that relate to these properties.[73–75] To aid in the further assessment of the electronic properties of these promising compounds, we have included the calculated full densities of states for these 13 promising compounds, which can be found online via Figshare. Overall, additional computational and experimental assessment of these physical properties for the promising compounds discovered here would be highly desirable.

**Table 2.** Summary of perovskite materials in material set 5A and 5B that have calculated HSE bandgaps within the range of 1.1-1.7 eV. There materials were sorted by their calculated PV efficiency values. Of these 28 (4) stable (metastable) materials in material set 5A and 5B, 12 (3) stable (metastable) materials pass the PV efficiency screening criterion, thus forming the set of most promising halide perovskite materials. Of the 12 (3) stable (metastable) materials that pass all screening criteria, to our knowledge 11 (1) of them have not been previously investigated. The material ID numbers provide a reference to the catalogued materials information in the spreadsheet included as part of the **SI**.

| **Stable materials** |
| --- |

| Material | ID # | $E_{hull}$ (meV/ atom) | PBE bandgap (eV) | Estimated HSE bandgap (eV) | Calculated HSE bandgap (eV) | PV efficiency (%) | Previously investigated? | Notes |
|---|---|---|---|---|---|---|---|---|
| $MA_{0.75}Cs_{0.25}SnI_3$ | 1695 | 0 | 1.17 | 1.70 | 1.20 | 24.9 | No | New compound |
| $MA_{0.875}Cs_{0.125}SnI_3$ | 1680 | 0.7 | 1.15 | 1.67 | 1.18 | 24.8 | No | New compound |
| $MA_{0.75}Rb_{0.25}SnI_3$ | 1740 | 0 | 1.17 | 1.69 | 1.21 | 24.6 | No | New compound |
| $KFeI_3$ | 1056 | 0 | 0.86 | 1.31 | 1.64 | 24.2 | No | New compound |
| $MAMnI_3$ | 647 | 0 | 0.82 | 1.25 | 1.27 | 23.9 | Experimentally, Ref. [60] and Ref. [76] Computationally, Ref. [46] | Spin-coated material was amorphous |
| $((CH_3)_2NH_2)Ag_{0.5}Bi_{0.5}I_3$ | 161 | 0 | 1.06 | 1.56 | 1.53 | 23.6 | No | New compound |
| $FA_{0.5}Cs_{0.5}SnI_3$ | 1800 | 0 | 1.09 | 1.60 | 1.11 | 23.4 | No | New compound |
| $CsMn_{0.875}Fe_{0.125}I_3$ | 272 | 0 | 0.91 | 1.38 | 1.33 | 23.3 | No | New compound |
| $FAAg_{0.5}Sb_{0.5}Br_3$ | 197 | 0 | 1.09 | 1.60 | 1.56 | 23.0 | No | New compound |
| $CsMn_{0.75}Co_{0.25}I_3$ | 270 | 0 | 0.93 | 1.40 | 1.50 | 22.8 | No | New compound |
| $MA_{0.5}Cs_{0.5}MnI_3$ | 1707 | 0 | 0.97 | 1.44 | 1.44 | 22.7 | No | New compound |
| $FA_{0.5}Rb_{0.5}SnI_3$ | 1845 | 0 | 1.12 | 1.64 | 1.12 | 22.7 | No | New compound |
| $CsBi_{0.5}Ag_{0.5}I_3$ | 17 | 12.7 | 1.11 | 1.63 | 1.55 | 22.4 | -- | -- |
| $Cs_{0.5}Na_{0.5}MnI_3$ | 259 | 7.4 | 1.06 | 1.56 | 1.58 | 21.9 | -- | -- |
| $RbMn_{0.75}Fe_{0.25}I_3$ | 356 | 0 | 0.90 | 1.37 | 1.34 | 21.9 | -- | -- |
| $RbMn_{0.875}Co_{0.125}I_3$ | 348 | 0 | 0.91 | 1.38 | 1.31 | 21.6 | -- | -- |
| $RbMn_{0.75}Co_{0.25}I_3$ | 351 | 0 | 0.94 | 1.41 | 1.31 | 20.8 | -- | -- |
| $RbMn_{0.875}Fe_{0.125}I_3$ | 353 | 0 | 0.91 | 1.37 | 1.35 | 20.6 | -- | -- |
| $RbMn_{0.5}Fe_{0.5}I_3$ | 355 | 0 | 0.88 | 1.33 | 1.46 | 20.6 | -- | -- |
| $Cs_{0.75}Na_{0.25}MnI_3$ | 260 | 0 | 0.93 | 1.39 | 1.33 | 20.4 | -- | -- |
| $Cs_{0.875}Rb_{0.125}MnI_3$ | 262 | 0 | 0.93 | 1.39 | 1.34 | 20.3 | -- | -- |
| $Cs_{0.875}Na_{0.125}MnI_3$ | 257 | 0 | 0.92 | 1.38 | 1.35 | 20.1 | -- | -- |
| $Cs_{0.75}Rb_{0.25}MnI_3$ | 265 | 0 | 0.93 | 1.39 | 1.34 | 20.0 | -- | -- |
| $RbMnI_3$ | 263 | 0 | 0.93 | 1.40 | 1.34 | 19.6 | -- | -- |
| $Cs_{0.5}Rb_{0.5}MnI_3$ | 264 | 0 | 0.93 | 1.40 | 1.34 | 19.6 | -- | -- |
| $Cs_{0.25}Rb_{0.75}MnI_3$ | 266 | 0 | 0.93 | 1.40 | 1.34 | 19.6 | -- | -- |
| $CsMnI_3$ | 1008 | 0 | 0.92 | 1.38 | 1.33 | 14.7 | -- | -- |
| $KMnI_3$ | 1068 | 10.0 | 0.95 | 1.42 | 1.33 | 14.3 | -- | -- |
| **Metastable materials** | | | | | | | | |

| Material | ID # | $E_{hull}$ (meV/atom) | PBE bandgap (eV) | Estimated HSE bandgap (eV) | Calculated HSE bandgap (eV) | PV efficiency (%) | Previously investigated? | Notes |
|---|---|---|---|---|---|---|---|---|
| $CsBi_{0.5}Cu_{0.5}I_3$ | 19 | 38 | 0.71 | 1.12 | 1.39 | 23.9 | Computationally, Ref. [48] | HSE bandgap of 1.3 eV |
| $((CH_3)_3C)GeI_3$ | 880 | 27.6 | 0.94 | 1.41 | 1.68 | 23.8 | No | New compound |
| $FASnI_3$ | 1317 | 32.6 | 1.13 | 1.65 | 1.23 | 22.9 | Experimentally and computationally, Refs. [77–79] | Well-known material |
| $RbAg_{0.5}Bi_{0.5}I_3$ | 65 | 32.1 | 1.11 | 1.63 | 1.53 | 22.6 | -- | -- |

## 2.6. Promising materials for silicon-perovskite tandem cells and quantum dot cells

As discussed in **Section 1**, we seek to not only suggest promising materials for use in only single junction solar cells, but also provide lists of promising materials for Si-perovskite tandem cells and quantum dot applications. Therefore, we have also screened the stable and metastable materials listed in **Table 1** (material sets 4A and 4B) that have calculated HSE bandgaps between 1.9-2.3 eV for the high bandgap material employed in Si-perovskite tandem cells and that have calculated HSE bandgaps between 0.8-1.4 eV for use in quantum dot applications. More information on the choice of these bandgap ranges is provided in **Section S2** of the **SI**. **Table 3** provides the list of promising stable and metastable materials for use in Si-perovskite tandem cells, while **Table 4** provides the analogous list for quantum dot applications. Overall, we have found 13 (26) materials that are promising for Si-perovskite tandem cell (quantum dot) applications, respectively.

**Table 3.** Summary of promising perovskite materials for Si-perovskite tandem solar cells that pass stability and calculated HSE bandgap screening criteria of materials with a bandgap between 1.9-2.3 eV. The material ID numbers provide a reference to the catalogued materials information in the spreadsheet included as part of the **SI**.

| Stable materials | | | | | |
|---|---|---|---|---|---|
| **Material** | **ID #** | **$E_{hull}$ (meV/atom)** | **Calculated HSE bandgap (eV)** | **Previously investigated?** | **Notes** |
| $MAFeI_3$ | 635 | 0 | 1.89 | No | New compound |
| $((CH_3)_2CH)BaI_3$ | 732 | 2.2 | 1.90 | No | New compound |
| $((CH_3)_2CH)CoI_3$ | 748 | 0 | 2.09 | No | New compound |
| $((CH_3)_2CH)SrI_3$ | 780 | 2.2 | 2.01 | No | New compound |
| $((CH_3)_2CH)ZnI_3$ | 784 | 12.6 | 2.14 | No | New compound |
| $((CH_3)_3C)CoI_3$ | 868 | 0 | 2.08 | No | New compound |
| $((CH_3)_3C)FeI_3$ | 876 | 0 | 2.03 | No | New compound |
| $((CH_3)_3C)MnI_3$ | 888 | 0 | 2.07 | No | New compound |
| $((CH_3)_3C)ZnI_3$ | 904 | 4.2 | 2.12 | No | New compound |
| **Metastable materials** | | | | | |
| **Material** | **ID #** | ***$E_{hull}$* (meV/atom)** | **Calculated HSE bandgap (eV)** | **Previously investigated?** | **Notes** |
| $RbCu_{0.5}Bi_{0.5}Cl_3$ | 59 | 45.1 | 2.23 | No | New compound |
| $RbMn_{0.5}Co_{0.5}I_3$ | 350 | 22.1 | 1.94 | No | New compound |
| $((CH_3)_2CH)CaI_3$ | 740 | 18.0 | 2.11 | No | New compound |
| $((CH_3)_3C)MgI_3$ | 884 | 17.6 | 2.19 | No | New compound |

**Table 4.** Summary of promising perovskite materials for quantum dot applications that pass stability and calculated HSE bandgap screening criteria of materials with a bandgap between 0.8-1.4 eV. The material ID numbers provide a reference to the catalogued materials information in the spreadsheet included as part of the SI.

| **Stable materials** | | | | | |
|---|---|---|---|---|---|
| **Material** | **ID #** | ***$E_{hull}$* (meV/atom)** | **Calculated HSE bandgap (eV)** | **Previously investigated?** | **Notes** |
| $Cs_{0.875}Na_{0.125}MnI_3$ | 257 | 0 | 1.35 | No | New compound |
| $Cs_{0.75}Na_{0.25}MnI_3$ | 260 | 0 | 1.33 | No | New compound |
| $Cs_{0.875}Rb_{0.125}Mn\ I_3$ | 262 | 0 | 1.34 | No | New compound |
| $RbMnI_3$ | 263 | 0 | 1.34 | No | New compound |
| $Cs_{0.5}Rb_{0.5}MnI_3$ | 264 | 0 | 1.34 | No | New compound |
| $Cs_{0.75}Rb_{0.25}MnI_3$ | 265 | 0 | 1.34 | No | New compound |
| $Rb_{0.75}Cs_{0.25}MnI_3$ | 266 | 0 | 1.34 | No | New compound |
| $CsMn_{0.875}Fe_{0.125}I_3$ | 272 | 0 | 1.33 | No | New compound |
| $RbMn_{0.875}Co_{0.125}I_3$ | 348 | 0 | 1.31 | No | New compound |
| $RbMn_{0.75}Co_{0.25}I_3$ | 351 | 0 | 1.31 | No | New compound |
| $RbMn_{0.875}Fe_{0.125}I_3$ | 353 | 0 | 1.35 | No | New compound |
| $RbMn_{0.75}Fe_{0.25}I_3$ | 356 | 0 | 1.34 | No | New compound |
| $MAMnI_3$ | 647 | 0 | 1.27 | Experimentally, Ref. [60] and Ref. [76] Computationally, Ref. [46] | Spin-coated material was amorphous |
| $CsMnI_3$ | 1008 | 0 | 1.33 | Experimentally, Ref. [80] | Hexagonal phase was fabricated |
| $KMnI_3$ | 1068 | 10.0 | 1.33 | No | New compound |
| $MA_{0.875}Cs_{0.125}SnI_3$ | 1680 | 0.7 | 1.18 | Experimentally, Ref. [81] | Successfully fabricated |
| $MA_{0.75}Cs_{0.25}SnI_3$ | 1695 | 0 | 1.20 | No | New compound |
| $MA_{0.5}Cs_{0.5}SnI_3$ | 1710 | 0 | 1.08 | No | New compound |
| $MA_{0.75}Rb_{0.25}SnI_3$ | 1740 | 0 | 1.21 | No | New compound |
| $MA_{0.5}Rb_{0.5}SnI_3$ | 1755 | 0 | 1.07 | No | New compound |
| $FA_{0.5}Cs_{0.5}SnI_3$ | 1800 | 0 | 1.11 | No | New compound |
| $FA_{0.5}Rb_{0.5}SnI_3$ | 1845 | 0 | 1.12 | No | New compound |
| **Metastable materials** | | | | | |
| **Material** | **ID #** | **$E_{hull}$ (meV/atom)** | **Calculated HSE bandgap (eV)** | **Previously investigated?** | **Notes** |
| $CsBi_{0.5}Cu_{0.5}I_3$ | 19 | 38.0 | 1.39 | Computationally, Ref. [76] | No bandgap value reported |
| $CsSb_{0.5}Ag_{0.5}I_3$ | 21 | 28.3 | 1.02 | Computationally, Ref. [82] | No bandgap value reported |
| $RbSb_{0.5}Ag_{0.5}I_3$ | 69 | 32.6 | 0.97 | Computationally, Ref. [82] | No bandgap value reported |
| $FASnI_3$ | 1317 | 32.6 | 1.23 | Experimentally and computationally, Refs. [77–79] | Well-known material |

From **Table 1** and **Table 4**, we found that several compounds containing Mn were predicted to be stable and have an optimal bandgap. This result is consistent with recent findings in other relevant materials. For example, Zhang *et al.* found that substituting Mn for Pb in $MAPbI_3$ resulted in enhanced stability of the perovskite.[76] In addition, Zou *et al.* found that doping Mn in place of Pb in $CsPbX_3$ ($X$ = Cl, Br, I) quantum dots increased the stability of the quantum dots relative to the pure Pb variant.[83] Additionally, the inclusion of Mn into $CsPbX_3$ ($X$ = Cl, Br, I) nanocrystals also induced favorable optical emission properties via Mn defect states, rendering Mn-doped $CsPbX_3$ nanocrystals intriguing materials for light emission applications like quantum dot light emitting diodes.[83–88] These results suggest that inclusion of transition metals into halide perovskites may aid in increasing the stability as well as tuning the bandgap, not only for bulk, single junction solar cells but also for quantum dot light emitting diodes. A caveat to the inclusion of transition metals like Mn is the possible presence of multiple redox states, which could create defect states within the bandgap, resulting in recombination centers that reduce the PV efficiency of the material. As the defect chemistry of these compounds is highly dependent on the composition and synthesis conditions, additional focused studies beyond the current work are required to further understand the potential of the transition metal-containing halide perovskites suggested here.

## 2.7. Qualitative chemical trends in bandgap and stability

In addition to identifying new candidate solar perovskite materials, it is valuable to gain improved understanding of what governs the bandgap and stability in halide perovskites. In this section, we have examined the general trend in our estimated HSE bandgaps and calculated stability for all materials simulated in this work. In the following discussion of **Figure 4(A-C)**, the data were partitioned into five groups based on their compositional alloying: the blue data "Organic pure" have only organic molecules on the *A*-site and unalloyed *B*- and *X*-sites; the green data "Organic alloyed" have only organic molecules on the *A*-site and alloyed *B*-sites; the red data "Inorganic pure" have only inorganic cations on the *A*-site and unalloyed *B*- and *X*-sites; the purple data "Inorganic alloyed" have only inorganic cations on the *A*-site and alloyed *B*- and/or *X*-sites; the black data "Organic/Inorganic alloyed" have an alloyed *A*-site with a mix of organic molecules and inorganic cations, while the *B*- and *X*-sites may be either unalloyed or alloyed.

**Figure 4(A-C)** qualitatively shows how the bandgap and stability of different groups of perovskites differ based on whether the *A*-, *B*-, or *X*-sites are alloyed. It is evident that all groups of materials scatter over a wide range. However, by examining the average values (indicated by the bold "X" symbols) in **Figure 4(A-C)**, trends in stability with alloying become apparent. First, in **Figure 4(A)**, the shift to lower $E_{hull}$ values between the blue ("Organic pure") and green ("Organic alloyed") bold "X" symbols indicates that, on average, the stability of a compound with an *A*-site fully occupied by organic molecules can be improved by alloying of the *B*-site. Next, in **Figure 4(B)**, the shift to lower $E_{hull}$ values between the red ("Inorganic pure") and purple ("Inorganic alloyed") bold "X" symbols also indicates that the stability of a compound with an *A*-site fully occupied by inorganic cations may be improved with alloying of the *B*- and *X*-sites. Finally, **Figure 4(C)** shows that the average $E_{hull}$ values indicated by the black bold "X" symbol for the "Organic/inorganic alloyed" material set results in values of $E_{hull}$ between the "Inorganic alloyed" and "Organic alloyed" material sets, and lower than the "Organic pure" and "Inorganic pure" material sets. In **Figure 4(D)**, we further explore the change in stability resulting from alloying by plotting the fraction of materials in each material set binned according to specific ranges in calculated $E_{hull}$ values. **Figure 4(D)** provides a quantitative representation of the distribution of $E_{hull}$ values for each material set. From the distributions in **Figure 4(D)**, the fraction of materials with $E_{hull}$ values in the most stable 0-50 meV/atom range is highest for the "Organic

alloyed” set (100%), followed by the “Organic/inorganic alloyed” set (90%), then the “Inorganic alloyed” set (70%), followed by the “Organic pure” set (60%), and finally the “Inorganic pure” set (20%). These trends suggest that doping the *A*-site with inorganic cations, and doping of the *B*- and *X*-sites of materials with organic cations on the *A*-site, plays a key role in stabilizing halide perovskite materials. These observed trends are consistent with a number of recent works which demonstrated that having mixed organic-inorganic species on the *A*-site can improve the stability of perovskite solar cells.[6–8,30,35–39]

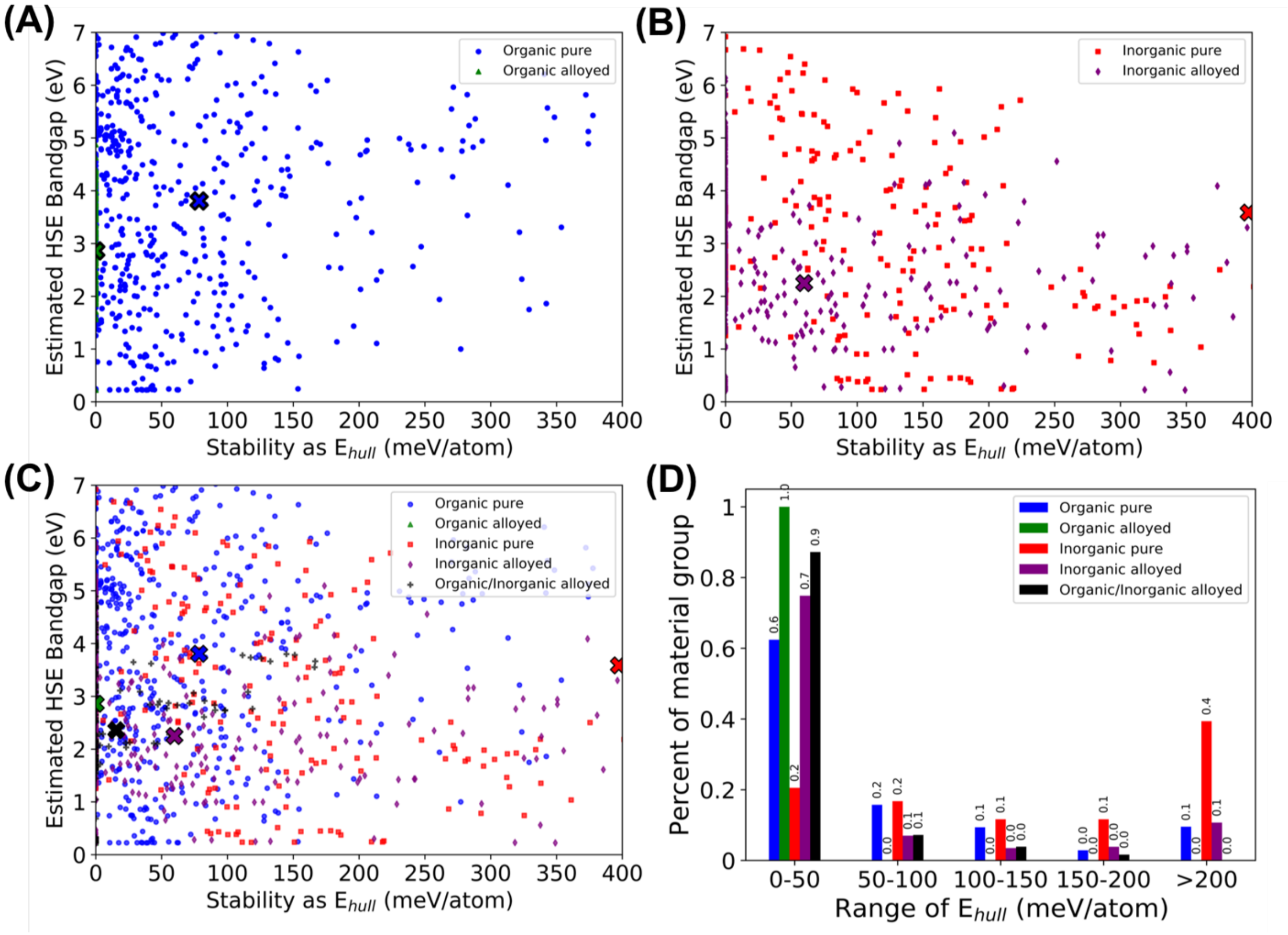

**Figure 4. (A-C)** Scatterplots of estimated HSE bandgap versus calculated stability for halide perovskites grouped into five categories, as described in the text. (A) “Organic pure” and “Organic alloyed” material sets, (B) “Inorganic pure” and “Inorganic alloyed” material sets, (C) the materials present in (A) and (B) along with the “Organic/Inorganic alloyed” material set. The colored bold “X” symbols denote average values of the stability and estimated HSE bandgap for the corresponding material set. (D) Distribution of calculated stabilities for each material set, given as a percentage of the materials comprising each respective material set as a function of different $E_{hull}$ ranges.

# 3. Summary and Conclusions

In this study, we have used high-throughput DFT methods to search for new halide perovskite materials that are comprised of non-toxic elements, are thermodynamically stable, and have an optimal bandgap for application in solar cells. By examining a total of 1845 simulated compounds, we found 720 (172) stable (metastable) compounds. Of the stable (metastable) compounds, 51 (11) compounds were predicted to have a bandgap within the optimal range for single junction solar cells. Of this pool of 62 compounds, 32 of them have calculated HSE bandgaps within the optimal range, and 15 of them were predicted to have high PV efficiencies in single junction solar cells. To our knowledge, 13 of these 15 materials have not been previously synthesized and tested (see **Table 2** for specific compositions). Some of these promising materials include $(CH_3NH_3)_{0.75}Cs_{0.25}SnI_3$, $((NH_2)_2CH)Ag_{0.5}Sb_{0.5}Br_3$, $CsMn_{0.875}Fe_{0.125}I_3$, $((CH_3)_2NH_2)Ag_{0.5}Bi_{0.5}I_3$, and $((NH_2)_2CH)_{0.5}Rb_{0.5}SnI_3$. From our set of stable and metastable compounds and estimated HSE bandgap values, we have also identified 13 materials that may be worth further investigation for Si-perovskite tandem cells (see **Table 3**) and 26 materials that may be worth further investigation for quantum dot applications (see **Table 4**). The large number of materials simulated in this study, together with other recent halide perovskite materials discovery studies, presents an opportunity to conduct data-driven research using machine learning tools to efficiently predict key properties of halide perovskites. Such data-driven studies, together with future detailed theoretical and experimental evaluations of the properties of the most promising compounds are highly desirable to flesh out which candidate compounds may be employed in the future generation of low cost, high efficiency solar photovoltaics.

# 4. Computational Methods

All Density Functional Theory (DFT) calculations were performed using the Vienna Ab initio Simulation Package (VASP).[89] To efficiently manage the large number of calculations in this study, the Materials Simulation Toolkit (MAST) was used.[51] The wave functions were modeled using a planewave basis set. The generalized gradient approximation (GGA) was used as the exchange and correlation functional. The pseudopotentials for each element type used the projector augmented wave (PAW) method[90] and were of the Perdew, Burke and Ernzerhof (PBE) type.[91] The valence electron configurations for each pseudopotential were those used by the Materials Project database, so that the thermodynamic phase stability of all compounds using Pymatgen can be performed accurately.[55,92] For compounds containing the transition metals Mn, Fe, Co, or Ni, the Hubbard $U$ correction (GGA+$U$) method was used.[93] The $U$-$J$ ($U_{eff}$) values were chosen to match those used by the Materials Project,[55,92] and were equal to 3.9, 5.3, 3.32, and 6.2 eV for Mn, Fe, Co and Ni, respectively. For our more quantitative bandgap calculations for the promising materials listed in **Table 1**, the hybrid functional of Heyd, Scuseria and Ernzerhof (HSE) was used, with a Hartree-Fock exchange fraction of 0.25 and inverse screening distance of 0.2 Å$^{-1}$.[94] For all calculations, the planewave cutoff energy was set to at least 400 eV, spin polarization was enabled, and the total energies were converged to 1 meV/cell. Additional details related to structural distortions and molecule orientation preference can be found in **Section S3** of the **SI**.

There were two main sets of perovskite materials modeled in this work: pure compounds (formula unit $ABX_3$) and alloyed compounds (formula unit $A_{1-x}A'_xB_{1-y}B'_yX_{3-z}X'_z$, where at least one of $x$, $y$, and $z$ is not equal to zero). For the pure compound set, a total of 960 materials were simulated using 1×1×1 (one formula unit) unit cells and an 8×8×8 (4×4×4) Monkhorst-Pack $k$-point mesh[95] for GGA (HSE, material set 5 only) calculations. For the alloyed compound set, a total of 882 materials were simulated using 2×2×2 (eight formula units) simulation cells and a 4×4×4 (2×2×2) Monkhorst-Pack $k$-point mesh for GGA (HSE, material set 5 only) calculations. For additional information on the full composition space of materials simulated in this work, see **Section S1** of the **SI**. While the unit cell parameters for all materials (both 1×1×1 single formula unit and 2×2×2 eight formula unit cells) were allowed to relax to non-cubic symmetries, no effort was made to model specific octahedral tilts which may occur in the true ground state structure. In particular, this means the 1×1×1 cells may not contain octahedral tilts which may manifest if larger unit cells were employed. These approximations were necessary to make the calculation of all compounds tractable. We regard our simulated structures to be a reasonable compromise between the ground state structure, which for some materials may be non-cubic, and the finite temperature cubic structure, which on average does not contain $BX_6$ octahedral tilts. We note here that previous DFT screening studies have also employed single formula unit supercells to represent diverse sets of halide perovskites, thus enacting similar approximations to the true ground state structure as used in this work.[13,47,71,96] In the following paragraphs, we remark on the expected scale of the effect on the bandgap and stability resulting from our use of single formula unit supercells.

Regarding the stability, the work of Young *et al.*[97] analyzed the change in energies for a set of representative inorganic halide perovskite compounds, and found that the energy change for different perovskite tilt systems (corresponding to, for example, a typical orthorhombic → cubic or tetragonal → cubic transformation) is about 5-25 meV/(formula unit), or about 1-5 meV/atom. This is a small difference in energy between competing structures, and is within typical DFT errors.

Regarding the differences in bandgap for different perovskite structures, we collected and analyzed DFT-calculated and experimental bandgaps from 41 different studies, spanning numerous materials and DFT calculation methods, and conducted tests on $KMnI_3$, $RbMnI_3$, $CsFeI_3$ and $KFeI_3$.[5,7,12,13,29,33,64,71,78,96,98–128] While this collection of 41 different studies is not an exhaustive search of the available literature of bandgaps of halide perovskites, we found this number of studies proved sufficient for us to estimate qualitative bandgap shifts between structural symmetries. Of these 41 studies examined, 23 of them performed DFT calculations and 18 of them performed experiments. A summary table of these findings is included in a spreadsheet as part of this **SI**, and is also publicly available on Figshare (see link in section summarizing the **Supplementary Information**). Materials for which experimental bandgap data for different structural symmetries were available include: $MAPbI_3$, $CsPbI_3$, $CsPbBr_3$, $CsPbCl_3$, and $CsSnI_3$. Materials for which DFT-calculated bandgap data for different structural symmetries were available include: $MAPbI_3$, $MAPbBr_3$, $MAPbCl_3$, $CsPbI_3$, $CsPbBr_3$, $FAPbI_3$, $CsSnI_3$, $KMnI_3$, $RbMnI_3$, $CsSnBr_3$, $CsFeI_3$, and $KFeI_3$. As a concrete example of the average approach we used consider the calculated bandgap difference of cubic and tetragonal $CsSnI_3$. Castelli *et al.*[117]calculated the bandgap of cubic (0.23 eV) and tetragonal (0.73 eV) $CsSnI_3$ while Huang *et al.*[114] also calculated the cubic (1.35 eV) and tetragonal (1.49 eV) bandgaps of $CsSnI_3$, but used quasiparticle GW methods. The average difference (i.e., the mean difference between $E_{gap}^{tetragonal} - E_{gap}^{cubic}$) of these calculated cubic and tetragonal bandgaps is 0.32 eV. In this example, and throughout this analysis, care was taken to only compare bandgaps calculated using the same DFT method (e.g. only comparing PBE, GW, etc.), and, when possible, the bandgaps of different symmetries for a particular material were compared within a single study. If no study that analyzed more than one symmetry was available for a particular material, we averaged the calculated bandgaps from the available studies which were calculated using the same DFT method. Therefore, the averages calculated from this method may also mix in effects from, for example, different DFT calculation settings used between studies, which may result in larger errors than if only comparisons of different structures within a particular study were examined. This method of averaging was necessary because very few studies contained bandgap information for multiple structures of the same composition. From analyzing the calculations of all different materials from these 23 different DFT studies as well as our own tests on $KMnI_3$, $RbMnI_3$, $CsFeI_3$ and $KFeI_3$, we found that the mean error in calculated bandgap between orthorhombic and cubic (tetragonal and cubic) materials was 0.17 (0.29) eV. From comparing the available experimental data of bandgaps for different structures from the remaining 18 studies we analyzed, the average difference between orthorhombic and cubic (tetragonal and cubic) materials was -0.04 (-0.05) eV. We note here that for the comparisons of experimental bandgaps between different symmetries, only data of $CsSnI_3$, $CsPbBr_3$, $CsPbI_3$ and $MAPbI_3$ in the orthorhombic and cubic phases was found and data of $CsPbCl_3$ and $MAPbI_3$ in the tetragonal and cubic phases was found, with a bias on data for $MAPbI_3$ (6 of the 18 experimental studies). Therefore, the average experimental bandgap shifts for different symmetries may be less robust than the average DFT bandgap shifts, as data for more materials was available from DFT studies. Overall, the expected differences in bandgap for a particular material based on different structural symmetries is within the established bandgap errors of the present work, especially for the case of known bandgap differences from experiment.

All phase stability calculations were conducted using the phase diagram analysis tools contained in the Pymatgen code package.[55] The space of possible decomposition products consist of all materials present in the Materials Project database, which, as of this writing, contains approximately 90,000 calculated inorganic compounds. To assess the stability of every halide perovskite material in the presence of water, we consider hydrogen in equilibrium with water vapor and oxygen gas, and have explicitly included hydrogen-containing compounds in the phase diagram calculations, even if the perovskite material being analyzed contained no hydrogen in its structure. We have used chemical potential values of $H_2$, $F_2$, $Br_2$, $Cl_2$ and $I_2$ consistent with ($T$, $P$) equal to $T$ = 298 K, $p(O_2)$ = 0.2 atm, a relative humidity of 30%, and the partial pressure of the halogen species set to $10^{-7}$ atm. This $T$, $P$, and relative humidity are approximate working conditions for perovskite solar cells and the halogen partial pressure is set to a somewhat arbitrary but representative low value. We have not considered a system also open to $O_2$ and $N_2$, as recent experiments analyzing the decomposition reactions of $MAPbI_3$, $MAPbBr_3$, $CsPbBr_3$, $CsPbI_3$, $CsSnI_3$ and $MASnI_3$ have found that the presence of water appears to have a catalytic effect, thus enhancing the decomposition kinetics of these compounds, as opposed to chemically reacting. The reaction with water could, in some cases, involve a dissolution reaction which could potentially provide a more stable decomposition pathway that explored here. However, we have not pursued this mechanism in this work. Thus, decomposition reactions of halide perovskites in the presence of water to form oxides or nitrides has not been observed and here we assume it is kinetically inhibited.[32,129–138] In addition, we note here that stability tests on a handful of halide perovskites when including a system open to $O_2$ and $N_2$ under air conditions resulted in very large instabilities (e.g. $E_{hull}$ values of approximately 1 eV/atom) with decomposition to various oxides. This observation regarding the thermodynamic instability of halide perovskites to oxide formation has also been recently pointed out in the work of Senocrate *et al.*,[139] where they similarly conclude that oxide formation appears to be kinetically inhibited.

Further, we have modified the space of possible decomposition products and stipulate that the organic species in the perovskite is stable against decomposition to other organics, including for the case of a loss of single H atom. For example, in $MAPbI_3$, $CH_3NH_3^+$ is not allowed to decompose to $CH_4$ and $NH_3$ or $CH_3NH_2$ and $H_2$. We believe not allowing the organic species to decompose to other organics is reasonable as numerous experimental investigations probing the decomposition reactions of $MAPbI_3$ have found, for example, that $CH_3NH_3^+$ tends to decompose to its precursor salt $CH_3NH_3I$ in the presence of water, and that further decomposition of CH3NH3+ to $CH_3NH_2$ and HI only occurs at elevated temperatures above room temperature. Further, we remark here that the present analysis is meant as a qualitative guide of bulk stability, and the precise experimental details of temperature, cell architecture, gas ambient, etc. ay result in different degradation kinetics or reaction pathways, the evaluation of which is outside the scope of the current work.[129,140,141] In addition, as there are few organic molecules and halide salts present in the Materials Project database, for each organic species considered in this work, we calculated its stable gaseous form (e.g. gaseous $CH_3NH_2$ when considering $CH_3NH_3^+$ in $MAPbI_3$) and its approximate precursor salt structure (e.g. solid $CH_3NH_3I$ when considering $CH_3NH_3^+$ in $MAPbI_3$). For the halide salt precursor structures, we used the existing MAI and FAI salt structures in Materials Project as starting structures. Finally, we note here that a similar method of stability screening was successfully used in a previous study by Jacobs *et al.* to screen perovskite oxides for stable, high activity solid oxide fuel cell cathodes, with qualitative stability values and trends that were found to agree well with experimental observations for a group of well-studied systems.[54] Additional details regarding the thermodynamic stability calculations can be found in **Section S4** of the **SI** and in the work of Jacobs *et al*.[54]

The PV efficiency of each promising material in **Table 2** was calculated using full dielectric functions obtained at the DFT-HSE level with the methods first developed by Shockley and Queisser,[140] modified to include contributions from reflection and incomplete absorption loss under the assumption of a normal incident light and a 0.5 μm-thick solar cell film. This value of 0.5 μm was chosen because it represents the typical perovskite layer thickness used in experimental assessment of perovskite-based solar cells.[5,8,9] Additional details on the PV efficiency calculations, including all equations necessary to perform the analysis, can be found in **Section S5** of the **SI**. Full data of the dielectric functions used to calculate the PV efficiency values are also included as part of the **SI**.

## Acknowledgements

R. Jacobs was supported by the National Science Foundation Software Infrastructure for Sustained Innovation (SI2) award #1148011 and G. Luo was supported by the Materials Research Science and Engineering Center (MRSEC) program, which is funded by the National Science Foundation Grant No. DMR1121288. Computing resources benefited from the University of Wisconsin-Madison MRSEC seed proposal program. Computational support was provided by the Extreme Science and Engineering Discovery Environment (XSEDE), which is supported by National Science Foundation Grant No. OCI-1053575. This research used computing resources of the National Energy Research Scientific Computing Center (NERSC), which is supported by the U.S. Department of Energy Office of Science. This research was also performed using the compute resources and assistance of the University of Wisconsin-Madison Center For High Throughput Computing (CHTC) in the Department of Computer Sciences.

## Supplementary Information

Supplementary information is available detailing the perovskite composition space explored, selection of elimination criteria used in screening, additional details for the computational methods, additional details for the thermodynamic stability calculations, and the PV efficiency criterion and calculation details. A spreadsheet is also included as part of the supplementary information, which provides the materials comprising each material set during screening and their associated stability and bandgap values. This data is also freely available on Figshare at (https://figshare.com/s/ef515baf994d7272215a). In addition, spreadsheets containing the data used to create each figure, as well as the halide perovskite bandgap data extracted from a series of computational and experimental studies is available. Finally, the key VASP input and output files for all materials examined in this work can also be found as part of the supplementary information.

## References

[1] D. Weber, *Zeitschrift fur Naturforsch. B* **1978**, *33b*.
[2] D. Weber, *Zeitschrift fur Naturforsch. B* **1978**, *33b*.
[3] A. Kojima, K. Teshima, Y. Shirai, T. Miyasaka, *J. Am. Chem. Soc.* **2009**, 6050.
[4] National Renewable Energy Lab (NREL): Best Research-Cell Efficiencies **2018**.
[5] G. E. Eperon, S. D. Stranks, C. Menelaou, M. B. Johnston, L. M. Herz, H. J. Snaith, *Energy Environ. Sci.* **2014**, *7*, 982.
[6] R. J. Sutton, G. E. Eperon, L. Miranda, E. S. Parrott, B. A. Kamino, J. B. Patel, M. T. Hörantner, M. B. Johnston, A. A. Haghighirad, D. T. Moore, H. J. Snaith, *Adv. Energy Mater.* **2016**, *6*, 1.
[7] J. H. Noh, S. H. Im, J. H. Heo, T. N. Mandal, S. Il Seok, *Nano Lett.* **2013**, *13*.
[8] D. P. Mcmeekin, G. Sadoughi, W. Rehman, G. E. Eperon, M. Saliba, M. T. Hörantner, A. Haghighirad, N. Sakai, L. Korte, B. Rech, M. B. Johnston, L. M. Herz, H. J. Snaith, *Science (80-. ).* **2016**, *351*, 3.
[9] S. D. Stranks, G. E. Eperon, G. Grancini, C. Menelaou, M. J. P. Alcocer, T. Leijtens, L. M. Herz, A. Petrozza, H. J. Snaith, *Science (80-. ).* **2013**, *342*.
[10] Q. Dong, Y. Fang, Y. Shao, P. Mulligan, J. Qiu, L. Cao, J. Huang, *Science (80-. ).* **2015**, *347*.
[11] G. Xing, N. Mathews, S. S. Lim, Y. M. Lam, S. Mhaisalkar, T. C. Sum, *Science (80-. ).* **2013**, *342*, 498.
[12] M. Du, *J. Phys. Chem. Lett.* **2015**, *6*.
[13] W. Yin, T. Shi, Y. Yan, W. Yin, T. Shi, Y. Yan, *Appl. Phys. Lett.* **2014**.
[14] D. Shi, V. Adinolfi, R. Comin, M. Yuan, E. Alarousu, A. Buin, Y. Chen, S. Hoogland, A. Rothenberger, K. Katsiev, Y. Losovyj, X. Zhang, P. A. Dowben, O. F. Mohammed, E. H. Sargent, O. M. Bakr, *Science (80-. ).* **2015**, *347*.
[15] C. S. Ponseca, T. J. Savenije, M. Abdellah, K. Zheng, A. Yartsev, T. Pascher, T. Harlang, P. Chabera, T. Pullerits, A. Stepanov, J. Wolf, V. Sundstro, *J. Am. Chem. Soc.* **2014**, *136*.
[16] F. Giustino, H. J. Snaith, *ACS Energy Lett.* **2016**, *1*.
[17] H. J. Snaith, *J. Phys. Chem. Lett.* **2013**, *4*.
[18] N. N. Lal, Y. Dkhissi, W. Li, Q. Hou, Y. Cheng, U. Bach, *Adv. Energy Mater.* **2017**, *7*, 1602761.
[19] M. T. Horantner, T. Leijtens, M. E. Ziffer, G. E. Eperon, M. G. Christoforo, M. D. McGehee, H. J. Snaith, *ACS Energy Lett.* **2017**, *2*.
[20] J. P. Mailoa, C. D. Bailie, E. C. Johlin, E. T. Hoke, A. J. Akey, W. H. Nguyen, D. Michael, J. P. Mailoa, C. D. Bailie, E. C. Johlin, E. T. Hoke, A. J. Akey, *Appl. Phys. Lett.*

**2015**, 121105.
[21] R. E. Beal, D. J. Slotcavage, T. Leijtens, A. R. Bowring, R. A. Belisle, W. H. Nguyen, G. F. Burkhard, E. T. Hoke, M. D. Mcgehee, *J. Phys. Chem. Lett.* **2016**, *7*, 2.
[22] S. Albrecht, M. Saliba, P. Correa, F. Lang, L. Korte, R. Schlatmann, M. K. Nazeeruddin, A. Hagfeldt, M. Gra, *Energy Environ. Sci.* **2016**, *9*, 81.
[23] J. Werner, C.-H. Weng, A. Walter, L. Fesquet, J. P. Seif, S. De Wolf, B. Niesen, C. Ballif, *J. Phys. Chem. Lett.* **2015**, *7*, 3.
[24] J. Werner, L. Barraud, A. Wlater, M. Brauninger, F. Sahli, D. Sacchetto, N. Tetreault, B. Paviet-salomon, S. Moon, C. Allebe, *ACS Energy Lett.* **2016**, *1*.
[25] Z. Tan, R. S. Moghaddam, M. L. Lai, P. Docampo, R. Higler, F. Deschler, M. Price, A. Sadhanala, L. M. Pazos, D. Credgington, F. Hanusch, T. Bein, H. J. Snaith, R. H. Friend, *Nat. Nanotechnol.* **2014**, *9*, 687.
[26] J. Song, J. Li, X. Li, L. Xu, Y. Dong, H. Zeng, *Adv. Mater.* **2015**, *27*, 7162.
[27] Y. Kim, E. Yassitepe, O. Voznyy, R. Comin, G. Walters, X. Gong, P. Kanjanaboos, A. F. Nogueira, E. H. Sargent, *ACS Appl. Mater. Interfaces* **2015**, *7*.
[28] Abhishek Swarnkar, A. R. Marshall, E. M. Sanehira, D. T. M. Boris D. Chernomordik, J. M. L. Jeffrey A. Christians, Tamoghna Chakrabarti, *Science (80-. ).* **2016**, *354*, 92.
[29] L. Protesescu, S. Yakunin, M. I. Bodnarchuk, F. Krieg, R. Caputo, C. H. Hendon, R. X. Yang, A. Walsh, M. V Kovalenko, *Nano Lett.* **2015**, *15*, 1.
[30] F. Hao, C. C. Stoumpos, D. H. Cao, R. P. H. Chang, M. G. Kanatzidis, *Nat. Photonics* **2014**, *8*, 489.
[31] Y. Rong, L. Liu, A. Mei, X. Li, H. Han, *Adv. Energy Mater.* **2015**, *5*, 1.
[32] T. Leijtens, G. E. Eperon, N. K. Noel, S. N. Habisreutinger, A. Petrozza, H. J. Snaith, *Adv. Energy Mater.* **2015**, *5*, 1.
[33] T. Krishnamoorthy, H. Ding, C. Yan, W. L. Leong, T. Baikie, Z. Zhang, S. Li, M. Asta, N. Mathews, S. G. Mhaisalkar, *J. Mater. Chem. A* **2015**, *3*.
[34] R. Ali, G. Hou, Z. Zhu, Q. Yan, Q. Zheng, *Chem. Mater.* **2018**, *30*.
[35] M. Saliba, T. Matsui, J.-Y. SEo, K. Domanski, E. Al., *Energy Environ. Sci.* **2016**, *9*, 1989.
[36] C. Yi, J. Luo, S. Meloni, A. Boziki, N. Ashari-Astani, C. Grätzel, S. M. Zakeeruddin, U. Röthlisberger, M. Grätzel, *Energy Environ. Sci.* **2016**, *9*, 656.
[37] J. Xiao, L. Liu, D. Zhang, N. De Marco, J. Lee, O. Lin, *Adv. Energy Mater.* **2017**, *7*, 1.
[38] Y. H. Park, I. Jeong, S. Bae, H. J. Son, P. Lee, J. Lee, C. Lee, M. J. Ko, *Adv. Funct. Mater.* **2017**, *27*, 21.
[39] J. Lee, D. Kim, H. Kim, S. Seo, S. M. Cho, N. Park, *Adv. Energy Mater.* **2015**, *5*.
[40] M. G. Ju, J. Dai, L. Ma, X. C. Zeng, *J. Am. Chem. Soc.* **2017**, *139*, 8038.
[41] L. K. Ono, E. J. Juarez-Perez, Y. Qi, *ACS Appl. Mater. Interfaces* **2017**, *9*, 30197.
[42] M. A. Green, T. Bein, *Nat. Mater.* **2015**, *14*, 559.
[43] J. Burschka, N. Pellet, S.-J. Moon, R. Baker-Humphry, P. Gao, M. Nzeeruddin, M. Gratzel, *Nature* **2013**, *499*, 4.
[44] K. Wojciechowski, T. Leijtens, S. Siprova, C. Schlueter, M. T. Hörantner, J. T. W. Wang, C. Z. Li, A. K. Y. Jen, T. L. Lee, H. J. Snaith, *J. Phys. Chem. Lett.* **2015**, *6*, 2399.
[45] M. R. Filip, F. Giustino, *J. Phys. Chem. C* **2016**, *120*.
[46] T. Nakajima, K. Sawada, *J. Phys. Chem. Lett.* **2017**, *8*.
[47] C. Kim, T. D. Huan, S. Krishnan, R. Ramprasad, *Sci. Data* **2017**, *4*, 1.
[48] G. Volonakis, M. R. Filip, A. A. Haghighirad, N. Sakai, B. Wenger, H. J. Snaith, F. Giustino, *J. Phys. Chem. Lett.* **2016**, *7*.

[49] Y. Cai, W. Xie, H. Ding, Y. Chen, K. Thirumal, L. H. Wong, *Chem. Mater.* **2017**, *29*.
[50] X. Zhao, J. Yang, Y. Fu, D. Yang, Q. Xu, L. Yu, S. Wei, L. Zhang, *J. Am. Chem. Soc.* **2017**, *139*.
[51] T. Mayeshiba, H. Wu, T. Angsten, A. Kaczmarowski, Z. Song, G. Jenness, W. Xie, D. Morgan, *Comput. Mater. Sci.* **2017**, *126*, 90.
[52] Y. Wu, P. Lazic, K. Persson, G. Ceder, *Energy Environ. Sci.* **2013**, *6*, 157.
[53] W. Sun, S. T. Dacek, S. P. Ong, G. Hautier, A. Jain, W. D. Richards, A. C. Gamst, K. A. Persson, G. Ceder, *Sci. Adv.* **2016**, *2*, 1.
[54] R. Jacobs, T. Mayeshiba, J. Booske, D. Morgan, *Adv. Energy Mater.* **2018**, *8*, 1.
[55] S. P. Ong, W. Davidson, A. Jain, G. Hautier, M. Kocher, S. Cholia, D. Gunter, V. L. Chevrier, K. A. Persson, G. Ceder, *Comput. Mater. Sci.* **2013**, *68*, 314.
[56] G. Pilania, J. E. Gubernatis, T. Lookman, *Comput. Mater. Sci.* **2017**, *129*, 156.
[57] J. Heyd, J. E. Peralta, G. E. Scuseria, R. L. Martin, *J. Chem. Phys.* **2005**, *123*, 174101.
[58] J. Heyd, G. E. Scuseria, *J. Chem. Phys.* **2004**, *121*, 1187.
[59] X. Hai, J. Tahir-Kheli, W. A. Goddard, *J. Phys. Chem. Lett.* **2011**, *2*, 212.
[60] T. M. Henderson, J. Paier, G. E. Scuseria, *Phys. Status Solidi B* **2011**, *248*, 767.
[61] C. H. Henry, *J. Appl. Phys.* **1980**, *51*, 4494.
[62] S. P. Bremner, M. Y. Levy, C. B. Honsberg, *Prog. Photovoltaics Res. Appl.* **2008**, *16*, 225.
[63] J. M. Ball, A. Petrozza, *Nat. Energy* **2016**, *1*.
[64] J. Kim, S. Lee, J. H. Lee, K. Hong, *J. Phys. Chem. Lett.* **2014**, *5*.
[65] I. Chung, J. H. Song, J. Im, J. Androulakis, C. D. Malliakas, H. Li, A. J. Freeman, J. T. Kenney, M. G. Kanatzidis, *J. Am. Chem. Soc.* **2012**, *134*, 8579.
[66] P. Xu, S. Chen, H. J. Xiang, X. G. Gong, S. H. Wei, *Chem. Mater.* **2014**, *26*, 6068.
[67] Y. Li, C. Zhang, X. Zhang, D. Huang, Q. Shen, Y. Cheng, W. Huang, *Appl. Phys. Lett.* **2017**, *111*, 1.
[68] J. Kang, L. W. Wang, *J. Phys. Chem. Lett.* **2017**, *8*, 489.
[69] S. Chakraborty, W. Xie, N. Mathews, M. Sherburne, R. Ahuja, M. Asta, S. G. Mhaisalkar, *ACS Energy Lett.* **2017**, *2*, 837.
[70] C. Wehrenfennig, G. E. Eperon, M. B. Johnston, H. J. Snaith, L. M. Herz, *Adv. Mater.* **2014**, *26*, 1584.
[71] C. Motta, F. El-mellouhi, S. Kais, N. Tabet, F. Alharbi, S. Sanvito, *Nat. Commun.* **2015**, *6*, 1.
[72] V. D. Innocenzo, G. Grancini, M. J. P. Alcocer, A. Ram, S. Kandada, S. D. Stranks, M. M. Lee, G. Lanzani, H. J. Snaith, A. Petrozza, *Nat. Commun.* **2014**, 1.
[73] Z. Xiao, Y. Yan, *Adv. Energy Mater.* **2017**, *7*, 1.
[74] W.-J. Yin, J.-H. Yang, J. Kang, Y. Yan, S.-H. Wei, *J. Mater. Chem. A* **2015**, *3*.
[75] W. J. Yin, T. Shi, Y. Yan, *J. Phys. Chem. C* **2015**, *119*, 5253.
[76] X. Zhang, J. Yin, Z. Nie, Q. Zhang, N. Sui, B. Chen, Y. Zhang, K. Qu, J. Zhao, H. Zhou, *RSC Adv.* **2017**, *7*, 37419.
[77] W. Liao, D. Zhao, Y. Yu, C. R. Grice, C. Wang, A. J. Cimaroli, P. Schulz, W. Meng, K. Zhu, R. G. Xiong, Y. Yan, *Adv. Mater.* **2016**, *28*, 9333.
[78] T. M. Koh, T. Krishnamoorthy, N. Yantara, C. Shi, W. L. Leong, P. P. Boix, A. C. Grimsdale, S. G. Mhaisalkar, N. Mathews, *J. Mater. Chem. A* **2015**, *3*, 14996.
[79] S. J. Lee, S. S. Shin, Y. C. Kim, D. Kim, T. K. Ahn, J. H. Noh, J. Seo, S. Il Seok, *J. Am. Chem. Soc.* **2016**, *138*, 3974.
[80] H. J. Seifert, K. H. Kischka, *Thermochim. Acta* **1978**, *27*.

[81] X. Liu, Z. Yang, C. C. Chueh, A. Rajagopal, S. T. Williams, Y. Sun, A. K. Y. Jen, *J. Mater. Chem. A* **2016**, *4*, 17939.
[82] J. Kangsabanik, V. Sugathan, A. Yadav, A. Yella, A. Alam, *Phys. Rev. Mater.* **2018**, *2*, 1.
[83] S. Zou, Y. Liu, J. Li, C. Liu, R. Feng, F. Jiang, Y. Li, J. Song, H. Zeng, M. Hong, X. Chen, *J. Am. Chem. Soc.* **2017**, *139*.
[84] M. D. Sampson, J. S. Park, R. D. Schaller, M. K. Y. Chan, A. B. F. Martinson, *J. Mater. Chem. A* **2017**, *5*, 3578.
[85] X. Yuan, S. Ji, M. C. De Siena, L. Fei, Z. Zhao, Y. Wang, H. Li, J. Zhao, D. R. Gamelin, *Chem. Mater.* **2017**, *29*, 8003.
[86] A. K. Guria, S. K. Dutta, S. Das Adhikari, N. Pradhan, *ACS Energy Lett.* **2017**, *2*, 1014.
[87] A. Swarnkar, V. K. Ravi, A. Nag, *ACS Energy Lett.* **2017**, *2*, 1089.
[88] W. Liu, Q. Lin, H. Li, K. Wu, I. Robel, J. M. Pietryga, V. I. Klimov, *J. Am. Chem. Soc.* **2016**, *138*, 14954.
[89] G. Kresse, J. Furthmüller, *Phys. Rev. B* **1996**, *54*, 11169.
[90] G. Kresse, D. Joubert, *Phys. Rev. B* **1999**, *59*, 1758.
[91] J. Perdew, K. Burke, M. Ernzerhof, *Phys. Rev. Lett.* **1996**, *77*, 3865.
[92] A. Jain, S. P. Ong, G. Hautier, W. Chen, W. D. Richards, S. Dacek, S. Cholia, D. Gunter, D. Skinner, G. Ceder, K. A. Persson, *APL Mater.* **2013**, *1*, 11002.
[93] J. Hubbard, *Proc. R. Soc. A Math. Phys. Eng. Sci.* **1963**, *276*, 238.
[94] J. Heyd, G. E. Scuseria, M. Ernzerhof, *J. Chem. Phys.* **2003**, *118*, 8207.
[95] H. Monkhorst, J. Pack, *Phys. Rev. B* **1976**, *13*, 5188.
[96] F. Brivio, A. B. Walker, A. Walsh, *APL Mater.* **2013**, *1*, 2.
[97] J. Young, J. M. Rondinelli, *J. Phys. Chem. Lett.* **2016**, *7*, 918.
[98] I. B. K. G. C. Papavassiliou, *Synth. Met.* **1995**, *71*.
[99] H. J. S. K. Heidrich, W. Schäfer, M. Schreiber, J. Söchtig, G. Trendel, J. Treusch, T. Grandke, *Phys. Rev. B* **1981**, *24*.
[100] J. K. S. D. M. Jang, K. Park, D. H. Kim, J. Park, F. Shojaei, H. S. Kang, J.-P. Ahn, J. W. Lee, *Nano Lett.* **2015**, *15*.
[101] M. G. K. C. C. Stoumpos, L. Frazer, D. J. Clark, Y. S. Kim, S. H. Rhim, A. J. Freeman, J. B. Ketterson, J. I. Jang, *J. Am. Chem. Soc.* **2015**, *137*.
[102] P. Y. D. Zhang, S. W. Eaton, Y. Yu, L. Dou, *J. Am. Chem. Soc.* **2015**, *137*.
[103] H. J. S. G. E. Eperon, G. M. Paterno, R. J. Sutton, A. Zampetti, A. A. Haghighirad, F. Cacialli, *J. Mater. Chem. A* **2015**, *3*.
[104] Z. Chen, C. Yu, K. Shum, J. J. Wang, W. Pfenninger, N. Vockic, J. Midgley, J. T. Kenney, *J. Lumin.* **2012**, *132*, 345.
[105] J. D. D. S. J. Clark, C. D. Flint, *J. Phys. Chem. Solids* **1981**, *42*.
[106] R. H. F. A. Sadhanala, F. Deschler, T. H. Thomas, S. E. Dutton, K. C. Goedel, F. C. Hanusch, M. L. Lai, U. Steiner, T. Bein, P. Docampo, D. Cahen, *J. Phys. Chem. Lett.* **2014**, *5*.
[107] M. G. K. C. C. Stoumpos, C. D. Malliakas, J. A. Peters, Z. Liu, M. Sebastian, J. Im, T. C. Chasapis, A. C. Wibowo, D. Y. Chung, A. J. Freeman, B. W. Wessels, *Cryst. Growth Des.* **2013**, *13*.
[108] B. W. W. M. Sebastian, J. A. Peters, C. C. Stoumpos, J. Im, S. S. Kostina, Z. Liu, M. G. Kanatzidis, A. J. Freeman, *Phys. Rev. B* **2015**, *92*.
[109] M. G. K. C. C. Stoumpos, C. D. Malliakas, *Inorg. Chem.* **2013**, *52*.
[110] K. S. C. Yu, Z. Chen, J. J. Wang, W. Pfenninger, N. Vockic, J. T. Kenney, *J. Appl. Phys.*

**2011**, *110*.
[111] Y. H. J. T. Li-Chuan, C. Chen-Shiung, T. Li-Chuan, *J. Phys. Condens. Matter* **2000**, *12*.
[112] I. A. M. Ahmad, G. Rehman, L. Ali, M. Shafiq, R. Iqbal, R. Ahmad, T. Khan, S. Jalali-Asadabadi, M. Maqbool, *J. Alloys Compd.* **2017**, *705*.
[113] I. A. G. Murtaza, *Phys. B Condens. Matter* **2011**, *406*.
[114] W. R. L. L.-y. Huang, Lambrecht, *Phys. Rev. B* **2013**, *88*.
[115] W. R. L. L.-y. Huang, Lambrecht, *Phys. Rev. B* **2016**, *93*.
[116] F. D. A. E. Mosconi, A. Amat, M. K. Nazeeruddin, M. Grätzel, *J. Phys. Chem. C* **2013**, *117*.
[117] K. W. J. I. E. Castelli, J. M. Garcia-Lastra, K. S. Thygesen, *APL Mater.* **2014**, *2*.
[118] F. D. A. P. Umari, E. Mosconi, *Sci. Rep.* **2014**, *4*.
[119] H. Z. Y. Wang, T. Gould, J. F. Dobson, H. Zhang, H. Yang, X. Yao, *Phys. Chem. Chem. Phys.* **2014**, *16*.
[120] C. K. J. Even, L. Pedesseau, J.-M. Jancu, *J. Phys. Chem. Lett.* **2013**, *4*.
[121] Y. T. J. Haruyama, K. Sodeyama, L. Han, *J. Phys. Chem. Lett.* **2014**, *5*.
[122] H. Z. Y. Wang, B. G. Sumpter, J. Huang, H. Zhang, P. Liu, H. Yang, *J. Phys. Chem. C* **2015**, *119*.
[123] A. W. K. T. Butler, J. M. Frost, *Mater. Horizons* **2015**, *2*.
[124] B. X. J. Feng, *J. Phys. Chem. Lett.* **2014**, *5*.
[125] K. Y. G. Giorgi, J.-I. Fujisawa, H. Segawa, *J. Phys. Chem. Lett.* **2013**, *4*.
[126] C. K. J. Even, L. Pedesseau, J.-M. Jancu, *Phys. status solidi – Rapid Res. Lett.* **2014**, *8*.
[127] A. M. R. S. Liu, F. Zheng, N. Z. Koocher, H. Takenaka, F. Wang, *J. Phys. Chem. Lett.* **2015**, *6*.
[128] F. D. A. A. Amat, E. Mosconi, E. Ronca, C. Quarti, P. Umari, M. K. Nazeeruddin, M. Gratzel, *Nano Lett.* **2014**, *14*.
[129] Z. Song, E. Al., *Adv. Energy Mater.* **2016**, *6*.
[130] A. Dualeh, E. Al., *Chem. Mater.* **2014**, *26*.
[131] T. A. Berhe, E. Al., *Energy Environ. Sci.* **2016**, *9*.
[132] G. Niu, et al., *J. Mater. Chem. A* **2014**, *2*.
[133] T. Burwig, E. Al., *J. Phys. Chem. Lett.* **2018**, *9*.
[134] A. Akbulatov, E. Al., *J. Phys. Chem. Lett.* **2017**, *8*.
[135] M. Kulbak, *J. Phys. Chem. Lett.* **2016**, *7*.
[136] J. Yang, T. Kelly, *Inorg. Chem.* **2017**.
[137] L. Dimesso, et al., *Mater. Chem. Phys.* **2017**, *197*.
[138] D. Mitzi, K. Liang, *J. Solid State Chem.* **1997**, *134*.
[139] J. M. Alessandro Senocrate, Tolga Acartürk, Gee Yeong Kim, Rotraut Merkle, Ulrich Starke, Michael Grätzel, *J. Mater. Chem. A* **2018**, *6*.
[140] H. J. Q. W. Shockley, *J. Appl. Phys.* **1961**, *32*.

# Supplementary Information:

## Materials discovery of stable and non-toxic halide perovskite materials for high-efficiency solar cells

Ryan Jacobs[1], Guangfu Luo[1,2], Dane Morgan[1,*]

[1] Department of Materials Science and Engineering, University of Wisconsin-Madison, Madison, WI 53706, USA

[2] Department of Materials Science and Engineering, Southern University of Science and Technology, Shenzhen 518055, P. R. China

*Corresponding author e-mail: ddmorgan@wisc.edu

## S1. Composition space explored

In this study, a number of pure and alloyed inorganic, organic, and mixed inorganic-organic halide perovskites were examined. Many of the alloying schemes summarized here were created as a result of those composition subspaces appearing promising for new solar cell materials based on analysis of an initial pool of pure materials. Below, we have briefly summarized the different compositional groupings of halide perovskite materials simulated in this work. All groups were used for the full starting data set. For a complete list of material compositions and associated calculated properties (composition, stability, PBE-level bandgap, estimated HSE bandgap), the reader is directed to the spreadsheet that accompanies this **SI**.

*Inorganic pure materials.* This set of unalloyed inorganic $ABX_3$ (i.e., only a single element present on each of the *A*-, *B*- and *X*-sites) compounds consisted of $A$ = {Li, Na, K, Rb, Cs}, $B$ = {Ba, Be, Ca, Cd, Co, Cu, Fe, Ge, Mg, Mn, Ni, Pb, Sn, Sr, Zn} and $X$ = {F, Cl, Br, I}, for a total of 300 materials.

*Organic pure materials.* This set of unalloyed organic $ABX_3$ (i.e., only a single organic molecule present on the *A*-site, and a single element present on the *B*- and *X*-sites) compounds consisted of $A$ = {$CH_3NH_3$, $CH_3CH_2NH_3$, $(NH_2)_2CH$, $(CH_3)_2NH_2$, $(CH_3)_3NH$, $(CH_3)_2CHNH_3$, $(CH_3)_2CH$, $(CH_3)_3C$, $CH_3CH_2CH_2$, $CH_3CH_2CHCH_3$, $CH_3CH_2$}, $B$ = {Ba, Be, Ca, Cd, Co, Cu, Fe, Ge, Mg, Mn, Ni, Pb, Sn, Sr, Zn} and $X$ = {F, Cl, Br, I}, for a total of 660 materials.

*Inorganic alloyed materials, group 1.* This set of alloyed inorganic materials began with $CsMnI_3$, $CsFeI_3$ and $CsCoI_3$ as host compounds. Then, Na and Rb were alloyed on the A-site of these three materials to form $(Na, Rb)_xCs_{1-x}(Mn, Fe, Co)I_3$, where $x$ = 0.125, 0.25, 0.5, 0.75 and 1. Separately for $CsMnI_3$, $CsFeI_3$ and $CsCoI_3$, the *X*-site was alloyed with Br and Cl to form $Cs(Mn, Fe, Co)I_{3(1-x)}(Br, Cl)_{3x}$, where again $x$ = 0.125, 0.25, 0.5, 0.75 and 1. A final alloying scheme is alloying only $CsMnI_3$ with Co and Fe on the B-site to create $CsMn_{1-x}(Fe, Co)_xI_3$, where here $x$ = 0.125, 0.25, 0.5 and 1. These alloying schemes combine for a total of 150 materials.

*Inorganic alloyed materials, group 2.* This set of inorganic materials consisted of $CsB_{0.5}B'_{0.5}X_{1.5}X'_{1.5}$ and $A_{0.5}A'_{0.5}B_{0.5}B'_{0.5}I_3$ as host compounds. For $CsB_{0.5}B'_{0.5}X_{1.5}X'_{1.5}$, $B$ = {Ba, Ca, Cd, Cu, Ge, Mg, Pb, Sn, Sr, Zn} and $X$ = {Cl, Br, I}. For $A_{0.5}A'_{0.5}B_{0.5}B'_{0.5}I_3$, $A$ = {K, Rb, Cs} and $B$ = {Ba, Ca, Cd, Cu, Ge, Mg, Pb, Sn, Sr, Zn}. These alloying schemes combine for a total of 336 materials.

*Inorganic double perovskite materials.* This set of inorganic double perovskite $A_2BB'X_6$ (i.e., $A$- and $X$-site are occupied by a single element and the $B$-site is occupied by two different elements in a face-centered cubic ordered arrangement) compounds consisted of $A$ = {K, Rb, Cs}, $B$, $B'$ = {Bi, Ag, Au, Cu, Tl, Sb} and $X$ = {Cl, Br, I} for a total of 72 materials.

*Organic double perovskite materials.* This set of organic double perovskite $A_2BB'X_6$ (i.e., $A$- and $X$-site are occupied by a single organic molecule and element, respectively, and the $B$-site is occupied by two different elements in a face-centered cubic ordered arrangement) compounds consisted of $A$ = {$CH_3NH_3$, $CH_3CH_2NH_3$, $(NH_2)_2CH$, $(CH_3)_2NH_2$, $(CH_3)_3NH$, $(CH_3)_2CHNH_3$}, $B$, $B'$ = {Bi, Ag, Au, Cu, Tl, Sb} and $X$ = {Cl, Br, I} for a total of 144 materials.

*Mixed inorganic/organic alloyed materials.* This set of alloyed compounds consists of $A_{1-x}A'_xBX_3$ ($x$ = 0.125, 0.25, 0.5), where $A$ is an organic cation and $A'$ is an inorganic element, and $A$ = {$CH_3NH_3$, $(NH_2)_2CH$}, $A'$ = {Rb, Cs}, $B$ = {Cu, Fe, Ge, Mn, Sn} and $X$ = {Cl, Br, I}. This inorganic/organic alloying scheme yielded 180 materials.

Overall, a total of 1845 halide perovskite materials were examined in this work.

**S2. Selection of elimination criteria**

*Toxicity*. We denoted an element as toxic if its precursor compounds that may likely be used in synthesis are known to be acutely toxic, according to information available from materials supplier websites like Sigma Aldrich. Of the elements we have included in our simulations, we denoted Pb, Cd and Be as toxic.

*Stability.* The stability of materials was determined using convex hull phase diagram analysis, which provides the stability as an energy above the convex hull ($E_{hull}$) of the phase diagram (see **Section S4** and the work of Jacobs *et al*. for additional details).[1] Following previous studies which have analyzed the predictive accuracy of material stability using DFT-based convex hull analysis compared to experimentally measured or observed materials stability and metastability,[1-4] we have denoted any materials that have an $E_{hull}$ value less than 15 meV/atom as predicted to be stable, which is the stability value of $MAPbI_3$. Further, since we were particularly interested in identifying new compounds that are more stable than the canonical lead-free halide perovskite material $MASnI_3$ (MA = methylammonium, $CH_3NH_3^+$), we have identified as metastable an additional set of materials which have $E_{hull}$ values greater than 15 meV/atom but less than 46 meV/atom, where this 46 meV/atom $E_{hull}$ value is the stability value for $MASnI_3$. These materials were denoted as metastable compounds as $MASnI_3$ has well-known stability issues, so these materials may also have some problems with stability. Finally, any materials that have calculated $E_{hull}$ values greater than 46 meV/atom were denoted as unstable.

*Bandgap.* For single junction solar cells, the ideal band gap value for a single junction solar cell is 1.34 eV, yielding the Shockley-Queisser efficiency limit of 33.68%.[5, 6] We approximated the desired bandgap within the range of 1.1-1.7 eV based on propagation of errors resulting from our fit of PBE vs. HSE bandgaps from previous studies (see **Figure S2**) and typical bandgap errors between HSE and experiment, as discussed in the main text. **Figure S1** below details the number of materials that pass the estimated HSE bandgap screening criterion as a function of the range of bandgaps considered. In addition to arising from a sensible propagation of errors calculation, our choice of the 1.1-1.7 eV bandgap range resulted in a computationally tractable number of HSE calculations. For Si-perovskite tandem solar cells, the ideal range of perovskite absorber bandgap to be paired with Si is within the range of 1.9-2.3 eV.[7-9] For quantum dot applications, the size confinement effects of quantum dots result in an increase of the bandgap of about 300 meV in the case of $CsPbI_3$ and $CsPbCl_3$ relative to bulk bandgap values.[10, 11] Assuming that this 300 meV bandgap increase for quantum dots is typical for halide perovskites as a whole, we shifted our single junction bandgap screening criterion down by 300 meV to search for promising materials for quantum dots in the bandgap range of 0.8-1.4 eV.

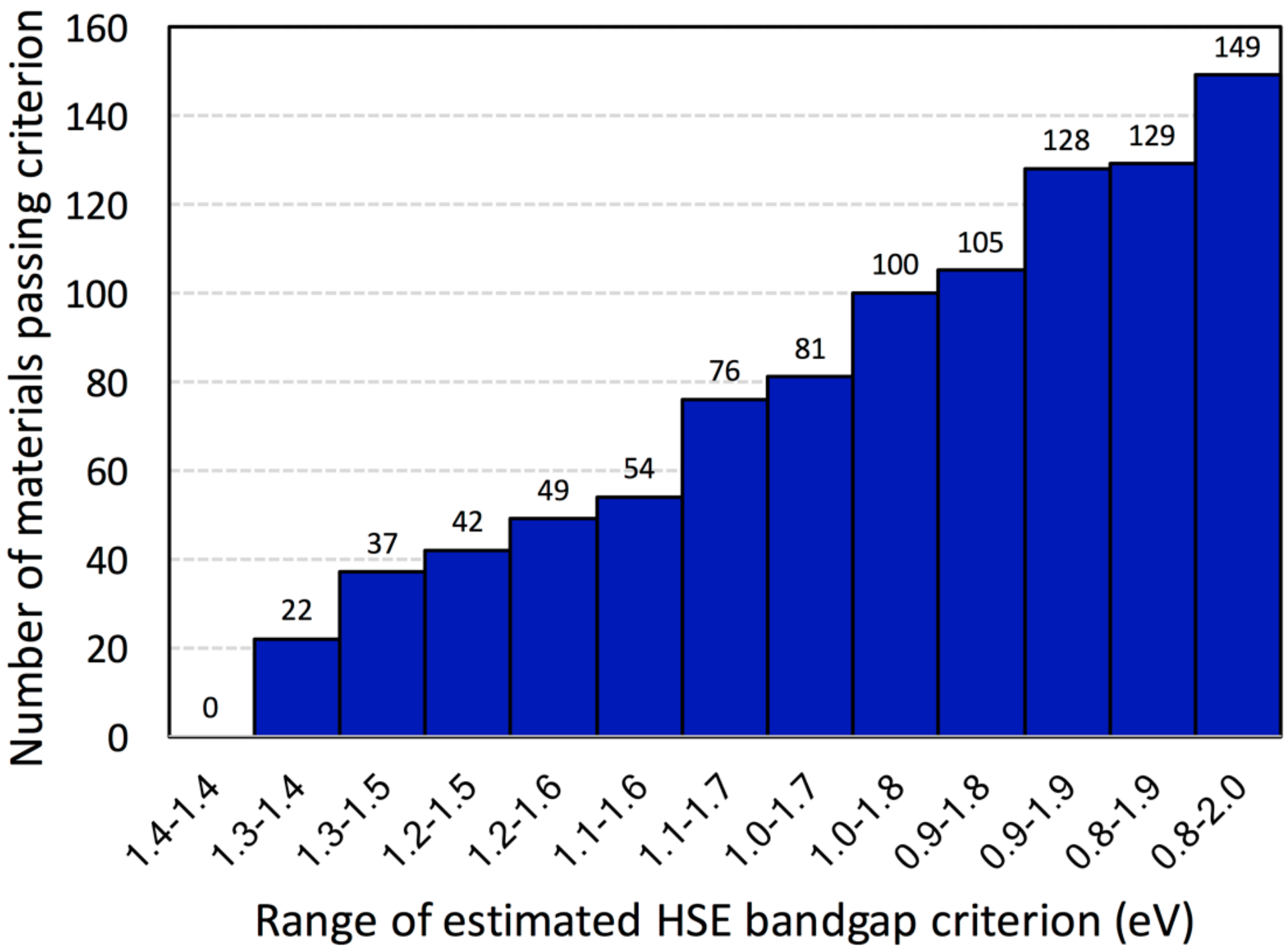

**Figure S1**. Number of perovskite materials passing the estimated HSE bandgap criterion as a function of the range of bandgaps considered. In this study, a range of 1.1-1.7 eV was chosen.

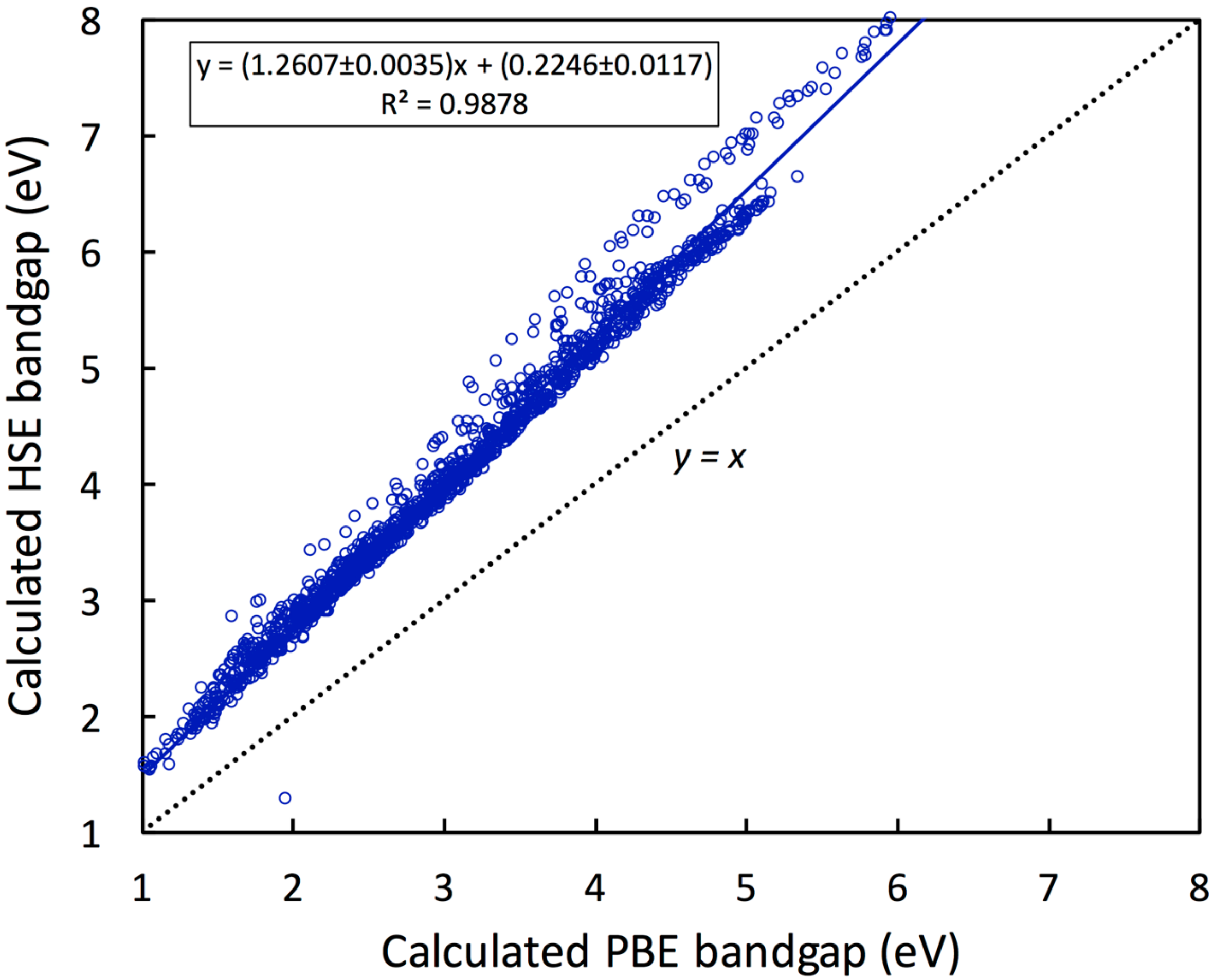

**Figure S2**. Calculated HSE bandgap versus calculated PBE bandgap for the perovskite materials simulated in the work of Kim, *et al.*[12] and Pilania, *et al.*[13] The black dashed line is the $y = x$ line. The line of best fit to the data produces the relationship for estimating the HSE bandgap from PBE bandgaps used in the main text, and is given as $E_{gap}^{HSE} = (1.2607 \pm 0.0035)E_{gap}^{PBE} + (0.2246 \pm 0.0117)$.

*Photovoltaic (PV) efficiency.* The optical absorption spectrum serves as a useful, higher-level property to screen against because the absorption spectrum and ideal single junction cell efficiency (which is calculable from the absorption spectrum)[6] of a candidate material can be compared to key solar absorbers like Si, GaAs and $MAPbI_3$.[14, 15] Given that the highest experimentally observed PV efficiency for hybrid perovskite solar cells is 22.7% at the time of this writing,[16] we chose the PV efficiency screening criterion to remove materials that have a PV efficiency of < 22.7%. The PV efficiency calculations were performed assuming a perovskite film thickness of 0.5 μm, which represents a typical perovskite layer thickness used in experimental assessment of perovskite-based solar cells.[17-19] We recognize that while this is a very stringent condition for determining whether a material may be a promising light absorber, the nature of the PV efficiency calculations conducted in this study are highly idealized, and thus represent an upper bound of the expected efficiency of the actual material in a real device.

**S3. Additional details for computational methods**

One challenge to our DFT modeling of a large number of halide perovskites with different chemistries and compositions is that the crystal structure of halide perovskites varies as a function of temperature, and the precise temperature that the perovskite assumes a cubic structure varies with the material composition.[20] For the specific case of $MAPbI_3$, the ground state structure is orthorhombic with tilted $PbI_6$ octahedra (space group $Pnma$), while the average tilt-free cubic phase (space group $Pm\bar{3}m$) is observed above 330.4 K.[21-23] In this work, for materials containing organic molecules on the $A$-site, the unrelaxed unit cell parameters were set to the values for cubic $MAPbI_3$, and for purely inorganic materials the unrelaxed unit cell parameters were set to the values for cubic $CsPbI_3$. All relaxations were performed with symmetry disabled to ensure materials could readily relax away from the initial constraint of cubic symmetry. While the unit cell parameters were allowed to relax to non-cubic symmetries, no effort was made to model specific octahedral tilts which may occur in the true ground state structure. These approximations were necessary so that calculation of all compounds was tractable. We regard our simulated structures to be a reasonable compromise between the ground state structure, which for some materials may be non-cubic, and the finite temperature cubic structure, which on average does not contain $BX_6$ octahedral tilts. We note here that previous DFT screening studies have also employed single formula unit supercells to represent diverse sets of halide perovskites, thus enacting similar approximations to the true ground state structure as used in this work.[12, 22-24] Here, we remark on the expected scale of the effect on the bandgap and stability resulting from our use of single formula unit supercells. Regarding the stability, the work of Young, *et al.*[25] has analyzed the change in energies for a set of representative inorganic halide perovskite compounds, and found that the energy change for different perovskite tilt systems (corresponding to, for example, a typical orthorhombic → cubic or tetragonal → cubic transformation) is about 5-25 meV/(formula unit), or about 1-5 meV/atom. This is a small difference in energy between competing structures, and is within typical DFT errors. Regarding the differences in bandgap for different perovskite structures, we collected and analyzed DFT-calculated and experimental bandgaps from 36 different studies, spanning numerous materials and DFT calculation methods.[11, 17, 21-24, 26-56] A summary table of these findings is included in a spreadsheet as part of this **SI**. From analyzing these 36 different studies, we found that the average calculated bandgap difference between orthorhombic and cubic (tetragonal and cubic) materials was 0.24 (0.23) eV. From comparing the available experimental data of bandgaps for different structures, the average difference between orthorhombic and cubic (tetragonal and cubic) materials was 0.12 (0.13) eV. Therefore, the expected differences in bandgap for a particular material based on different structural symmetries is within the established bandgap errors of the present work, especially for the case of known bandgap differences from experiment.

An additional challenge to modeling a large variety of these materials is specific to halide perovskites containing organic molecules on the *A*-site. The organic molecules may reside in different orientations, and polar molecules may be subject to dipole ordering. For the specific case of $MAPbI_3$, experimental investigations have shown that the MA molecules in $MAPbI_3$ rotate freely under ambient conditions, and that ferroelectric and antiferroelectric domains may form under different device biasing conditions.[57, 58] Previous DFT calculations have shown that the organic molecules in $MAPbI_3$ oriented along different crystallographic directions are nearly degenerate in energy.[22, 58] The work of Brivio*, et al.*[22] found a slight energetic preference of MA molecules in $MAPbI_3$ in the [100] orientation by about 15 meV/atom. Therefore, for materials containing organic molecules on the *A*-site, we simulated all materials with the dipole of the organic molecule oriented along the [100] direction.

This ferroelectric molecule ordering was used in order to generate a set of consistent and simplified structures to conduct high-throughput computational analysis across many chemical systems, most of which have not been experimentally investigated. Because of this, it is possible that some of the materials explored in this work were not modeled in their true ground state configuration. Therefore, while the values of bandgap, total energy, stability, etc. are not quantitatively accurate for all systems, we believe the qualitative trends across different chemistries and material compositions is preserved, which is sufficient for the present study of identifying specific chemical and composition spaces containing promising perovskite halide materials as discussed above. In addition to the issue of molecule orientation, for the case of $MAPbI_3$ we modeled $2\times2\times2$ (eight formula unit) simulation cells to test the effect of MA molecule ordering. We found that ferroelectric ordering of the MA molecules yielded the most stable structure, consistent with known observations of the spontaneous polarization present in these materials.[57-60] We also found that certain antiferroelectric arrangements were only about 5 meV/atom less stable than the ferroelectric arrangement and are thus competitive in energy. Overall, these tests indicate that our choice of ferroelectric molecule ordering along the [100] direction is reasonable for a high-throughput analysis of halide perovskites containing polar organic molecules.

Regarding the calculation of bandgaps, it is well-known that DFT underestimates the bandgap at the PBE level.[61] In order to keep the number of HSE calculations tractable, we used a linear scaling relationship between calculated PBE-level and HSE-level bandgaps to estimate the HSE bandgaps for all stable and metastable compounds using calculated PBE values, as discussed in the main text and in **Section S2**. For materials that passed the PBE and estimated HSE screening criteria, we explicitly calculated the HSE bandgaps. Previous studies analyzing the accuracy of HSE for bandgaps have found that mean absolute errors between HSE and experimental bandgaps are about 0.2-0.3 eV.[61-64]

### S4. Additional details for thermodynamic stability calculations

The high temperature chemical phase stability of all compounds screened in this study was analyzed using the multicomponent phase diagram modules contained within the Pymatgen toolkit (version 4.2.0).[65] More information on the specific Pymatgen modules, classes, and class methods used to conduct the phase stability analysis is provided in the work of Jacobs, *et al.*[1] It was assumed that every potential halide perovskite material would be subject to an environment that is open both to $H_2$ and the appropriate halogen species: $I_2$, $Br_2$, $F_2$ and $Cl_2$. The chemical potential of $H_2$ was set by the chemical potential of $O_2$ and equilibrium with $H_2O$ vapor at 298 K. We assumed a humid operating environment with a relative humidity (*RH*) of 30%, which is an approximate value for the amount of $H_2O$ present in ambient air, which is the working condition of terrestrial solar cells. To obtain the chemical potential of $H_2$ in equilibrium with $H_2O$, the chemical potential of $O_2$ was set such that the temperature was 298 K and the partial pressure of $O_2$ was 0.2 atm. Following the work of Jacobs *et al.*, the GGA/GGA+$U$ energy shifts were applied for halide perovskites modeled with GGA+$U$.[1]

The values of the chemical potentials for $O_2$, $H_2$, $F_2$, $Br_2$, $Cl_2$ and $I_2$ used in the phase diagram analysis tools in Pymatgen were derived from standard gas phase thermodynamics equations. As all solid phase DFT energies are under conditions of $T$ = 0 K and $P$ = 0 atm, the temperature and pressure values typical for solar cell operation ($T$ = 298 K, $p(O_2)$ = 0.2 atm and the partial pressure of all halides was set at $10^{-7}$ atm) are built into the gas chemical potentials. We chose these halogen chemical potentials to represent an expected low partial pressure, but not so low that they drive unphysical instability in the materials. We calculated the gas chemical potentials using experimental data from the NIST chemistry webbooks[66] and standard thermochemistry equations as detailed in previous works.[67, 68] The chemical potential values for O, H, F, Br, Cl and I were -5.25, -4.91, -1.85, -1.60, -1.77, -1.44 eV/atom, respectively.

## S5. PV efficiency criterion and PV efficiency calculation details

The PV power efficiency, $\eta$, of a planar p-n junction solar cell is calculated by considering the detailed balance limit, spectrum reflection, and incomplete light absorption due to finite thickness, as defined in Eqn. S1. The equations presented in this section to calculate the PV power efficiency were implemented in a Mathematica script which has been included as part of this **SI**. The well-known Shockley-Queisser efficiency limit[6] approximated the solar power density with black-body radiation and assumed zero reflectance and infinite thickness.

$$\eta = \frac{\int_{E_g}^{\infty} Sun(E)\frac{E_g}{E}\times\left(1-R(E)\right)\times\left(1-e^{-2\pi\alpha(E)L_0}\right)dE}{\int_0^{\infty} Sun(E)dE}\nu(T_c,E_g)m(T_c,E_g), \qquad \text{(S1)}$$

where $Sun(E)$ is the power density of solar spectrum ASTM G173-03 with consideration of a 37° tilting relative to horizontal (which is the result of sunlight entering the atmosphere), and $E_g$ is the band gap of a single-junction solar cell.[69] In this work, we employ the HSE band gap. The light reflectance under normal incidence (after the sun has entered the atmosphere), $R(E)$, and absorption coefficient, $\alpha(E)$, are defined in Eqn. S2 and S3, respectively, and depend on dielectric function $\varepsilon_1(E)$ and $\varepsilon_2(E)$ through Eqn. S4. The values of $n$, $k$ and $c$ in Eqn. S2, S3 and S4 are the refractive index, extinction coefficient, and vacuum speed of light, respectively. $L_0$ is the effective light absorption thickness, which is assumed to be 0.5 μm in our calculations, a value representative of the perovskite film thickness in numerous reports of experimental perovskite solar cell efficiencies.[17-19] The calculated dielectric functions for all materials in material set 5A and 5B are provided with the other main VASP calculation files as part of this **SI**.

$$R(E)=\frac{(n-1)^2+k^2}{(n+1)^2+k^2} \tag{S2}$$

$$\alpha(E)=\frac{2k}{c}E \tag{S3}$$

$$2nk=\varepsilon_2(E);\ n^2-k^2=\varepsilon_1(E) \tag{S4}$$

$v(T_c, E_g)$ and $m(T_c, E_g)$ are open-circuit voltage factor and impedance matching factor (or fill factor) as defined in Eqn. S5 and S6, respectively.

$$v(T_c,E_g)=\frac{k_B T_c}{E_g}z_{op}, \tag{S5}$$

$$m(T_c,E_g)=\frac{z_m^2}{\left(1+z_m-e^{-z_m}\right)\left(z_m+\ln(1+z_m)\right)}. \tag{S6}$$

$T_c$ is the working temperature of solar cell, which is assumed to be 300 K; $z_{op}$ and $z_m$ are defined by Eqn. S7 and S8, respectively; $W$ is the Lambert-W function.

$$z_{op}=\ln\left(\frac{\int_{E_g}^{\infty} Sun(E)/E\,dE}{\frac{4\pi}{h^3c^2}\int_{E_g}^{\infty} E^2/(e^{E/k_BT_c}-1)\,dE}\right), \tag{S7}$$

$$z_m=W\left(e^{1+z_{op}}\right)-1. \tag{S8}$$

Since our calculations of dielectric function, $\varepsilon_1(E)$ and $\varepsilon_2(E)$, did not include the excitonic effects or the electron-hole interactions, a minor error with the power efficiency $\eta$ could be expected. Our tests show that the power efficiency of bulk GaAs with and without excitonic effects differ only by 1%.

## References

[1] R. Jacobs, T. Mayeshiba, J. Booske, D. Morgan, *Advanced Energy Materials* **2018**, 1702708.

[2] W. Sun, S. T. Dacek, S. P. Ong, G. Hautier, A. Jain, W. D. Richards, A. C. Gamst, K. A. Persson, G. Ceder, *Science Advances* **2016**, 2.

[3] Y. Wu, P. Lazic, G. Hautier, K. Persson, G. Ceder, *Energ. Environ. Sci.* **2013**, 6, 157.

[4] J. E. Saal, S. Kirklin, M. Aykol, B. Meredig, C. Wolverton, *JOM* **2013**, 65, 1501.

[5] S. P. Bremner, M. Y. Levy, C. B. Honsberg, *Progress in Photovoltaics: Research and Applications* **2008**, 16, 225.
[6] W. Shockley, H. J. Queisser, *Journal of Applied Physics* **1961**, 32, 510.
[7] J. P. Mailoa, C. D. Bailie, E. C. Johlin, E. T. Hoke, A. J. Akey, W. H. Nguyen, M. D. McGehee, T. Buonassisi, *Applied Physics Letters* **2015**, 106, 121105.
[8] J. Werner, L. Barraud, A. Walter, M. Bräuninger, F. Sahli, D. Sacchetto, N. Tétreault, B. Paviet-Salomon, S.-J. Moon, C. Allebé, M. Despeisse, S. Nicolay, S. De Wolf, B. Niesen, C. Ballif, *ACS Energy Letters* **2016**, 1, 474.
[9] R. E. Beal, D. J. Slotcavage, T. Leijtens, A. R. Bowring, R. A. Belisle, W. H. Nguyen, G. F. Burkhard, E. T. Hoke, M. D. McGehee, *The Journal of Physical Chemistry Letters* **2016**, 7, 746.
[10] A. Swarnkar, A. R. Marshall, E. M. Sanehira, B. D. Chernomordik, D. T. Moore, J. A. Christians, T. Chakrabarti, J. M. Luther, *Science* **2016**, 354, 92.
[11] L. Protesescu, S. Yakunin, M. I. Bodnarchuk, F. Krieg, R. Caputo, C. H. Hendon, R. X. Yang, A. Walsh, M. V. Kovalenko, *Nano Letters* **2015**, 15, 3692.
[12] C. Kim, T. D. Huan, S. Krishnan, R. Ramprasad, *Scientific Data* **2017**, 4, 170057.
[13] G. Pilania, J. E. Gubernatis, T. Lookman, *Computational Materials Science* **2017**, 129, 156.
[14] M.-G. Ju, J. Dai, L. Ma, X. C. Zeng, *Journal of the American Chemical Society* **2017**, 139, 8038.
[15] X.-G. Zhao, J.-H. Yang, Y. Fu, D. Yang, Q. Xu, L. Yu, S.-H. Wei, L. Zhang, *Journal of the American Chemical Society* **2017**, 139, 2630.
[16] *National Renewable Energy Lab (NREL): Best Research-Cell Efficiencies:* https://www.nrel.gov/pv/assets/images/efficiency-chart.png **2017**.
[17] G. E. Eperon, S. D. Stranks, C. Menelaou, M. B. Johnston, L. M. Herz, H. J. Snaith, *Energ. Environ. Sci.* **2014**, 7, 982.
[18] D. P. McMeekin, G. Sadoughi, W. Rehman, G. E. Eperon, M. Saliba, M. T. Hörantner, A. Haghighirad, N. Sakai, L. Korte, B. Rech, M. B. Johnston, L. M. Herz, H. J. Snaith, *Science* **2016**, 351, 151.
[19] S. D. Stranks, G. E. Eperon, G. Grancini, C. Menelaou, M. J. P. Alcocer, T. Leijtens, L. M. Herz, A. Petrozza, H. J. Snaith, *Science* **2013**, 342, 341.
[20] T. Baikie, Y. Fang, J. M. Kadro, M. Schreyer, F. Wei, S. G. Mhaisalkar, M. Graetzel, T. J. White, *Journal of Materials Chemistry A* **2013**, 1, 5628.
[21] Y. Wang, T. Gould, J. F. Dobson, H. Zhang, H. Yang, X. Yao, H. Zhao, *Physical Chemistry Chemical Physics* **2014**, 16, 1424.
[22] F. Brivio, A. B. Walker, A. Walsh, *Apl Mater* **2013**, 1, 042111.
[23] C. Motta, F. El-Mellouhi, S. Kais, N. Tabet, F. Alharbi, S. Sanvito, *Nature Communications* **2015**, 6, 7026.
[24] W.-J. Yin, T. Shi, Y. Yan, *Applied Physics Letters* **2014**, 104, 063903.
[25] J. Young, Rondinelli, J. M., *Journal of Physcial Chemistry Letters* **2016**, 7, 918.
[26] E. Mosconi, A. Amat, M. K. Nazeeruddin, M. Grätzel, F. De Angelis, *The Journal of Physical Chemistry C* **2013**, 117, 13902.
[27] I. E. Castelli, J. M. Garcia-Lastra, K. S. Thygesen, K. W. Jacobsen, *Apl Mater* **2014**, 2, 081514.
[28] P. Umari, E. Mosconi, F. De Angelis, *Scientific Reports* **2014**, 4, 4467.
[29] J. Even, L. Pedesseau, J.-M. Jancu, C. Katan, *The Journal of Physical Chemistry Letters* **2013**, 4, 2999.

[30] J. Haruyama, K. Sodeyama, L. Han, Y. Tateyama, *The Journal of Physical Chemistry Letters* **2014**, 5, 2903.
[31] Y. Wang, B. G. Sumpter, J. Huang, H. Zhang, P. Liu, H. Yang, H. Zhao, *The Journal of Physical Chemistry C* **2015**, 119, 1136.
[32] K. T. Butler, J. M. Frost, A. Walsh, *Materials Horizons* **2015**, 2, 228.
[33] J. Kim, S.-H. Lee, J. H. Lee, K.-H. Hong, *The Journal of Physical Chemistry Letters* **2014**, 5, 1312.
[34] T. Krishnamoorthy, H. Ding, C. Yan, W. L. Leong, T. Baikie, Z. Zhang, M. Sherburne, S. Li, M. Asta, N. Mathews, S. G. Mhaisalkar, *Journal of Materials Chemistry A* **2015**, 3, 23829.
[35] M.-H. Du, *The Journal of Physical Chemistry Letters* **2015**, 6, 1461.
[36] J. Feng, B. Xiao, *The Journal of Physical Chemistry Letters* **2014**, 5, 1278.
[37] G. Giorgi, J.-I. Fujisawa, H. Segawa, K. Yamashita, *The Journal of Physical Chemistry Letters* **2013**, 4, 4213.
[38] C. Eames, J. M. Frost, P. R. F. Barnes, B. C. O'Regan, A. Walsh, M. S. Islam, *Nature Communications* **2015**, 6, 7497.
[39] J. Even, L. Pedesseau, J.-M. Jancu, C. Katan, *physica status solidi (RRL) – Rapid Research Letters* **2014**, 8, 31.
[40] J. M. Azpiroz, E. Mosconi, J. Bisquert, F. De Angelis, *Energy & Environmental Science* **2015**, 8, 2118.
[41] M. R. Filip, S. Hillman, A. A. Haghighirad, H. J. Snaith, F. Giustino, *The Journal of Physical Chemistry Letters* **2016**, 7, 2579.
[42] S. Liu, F. Zheng, N. Z. Koocher, H. Takenaka, F. Wang, A. M. Rappe, *The Journal of Physical Chemistry Letters* **2015**, 6, 693.
[43] G. Volonakis, Filip, M. R., Haghighirad, A. A., Sakai, N., Wenger, B., Snaith, H. J., Giustino, F., *Journal of Physical Chemistry Letters* **2016**, 7, 1254.
[44] M. R. Filip, F. Giustino, *The Journal of Physical Chemistry C* **2016**, 120, 166.
[45] M. R. Filip, G. E. Eperon, H. J. Snaith, F. Giustino, *Nature Communications* **2014**, 5, 5757.
[46] A. Amat, E. Mosconi, E. Ronca, C. Quarti, P. Umari, M. K. Nazeeruddin, M. Gratzel, F. De Angelis, *Nano Letters* **2014**, 14, 3608.
[47] C. K. MØLler, *Nature* **1958**, 182, 1436.
[48] C. C. Stoumpos, C. D. Malliakas, J. A. Peters, Z. Liu, M. Sebastian, J. Im, T. C. Chasapis, A. C. Wibowo, D. Y. Chung, A. J. Freeman, B. W. Wessels, M. G. Kanatzidis, *Crystal Growth & Design* **2013**, 13, 2722.
[49] M. Sebastian, J. A. Peters, C. C. Stoumpos, J. Im, S. S. Kostina, Z. Liu, M. G. Kanatzidis, A. J. Freeman, B. W. Wessels, *Physical Review B* **2015**, 92, 235210.
[50] C. C. Stoumpos, C. D. Malliakas, M. G. Kanatzidis, *Inorganic Chemistry* **2013**, 52, 9019.
[51] C. Yu, Z. Chen, J. J. Wang, W. Pfenninger, N. Vockic, J. T. Kenney, K. Shum, *Journal of Applied Physics* **2011**, 110, 063526.
[52] L.-y. Huang, W. R. L. Lambrecht, *Physical Review B* **2013**, 88, 165203.
[53] T. Li-Chuan, C. Chen-Shiung, T. Li-Chuan, Y. H. Jung, *Journal of Physics: Condensed Matter* **2000**, 12, 9129.
[54] L.-y. Huang, W. R. L. Lambrecht, *Physical Review B* **2016**, 93, 195211.
[55] M. Saliba, T. Matsui, J.-Y. Seo, K. Domanski, J.-P. Correa-Baena, M. K. Nazeeruddin, S. M. Zakeeruddin, W. Tress, A. Abate, A. Hagfeldt, M. Gratzel, *Energy & Environmental Science* **2016**, 9, 1989.

[56] T. M. Koh, T. Krishnamoorthy, N. Yantara, C. Shi, W. L. Leong, P. P. Boix, A. C. Grimsdale, S. G. Mhaisalkar, N. Mathews, *Journal of Materials Chemistry A* **2015**, 3, 14996.
[57] A. M. A. Leguy, J. M. Frost, A. P. McMahon, V. G. Sakai, W. Kockelmann, C. Law, X. Li, F. Foglia, A. Walsh, B. C. O'Regan, J. Nelson, J. T. Cabral, P. R. F. Barnes, *Nature Communications* **2015**, 6, 7124.
[58] W.-J. Yin, J.-H. Yang, J. Kang, Y. Yan, S.-H. Wei, *Journal of Materials Chemistry A* **2015**, 3, 8926.
[59] J. M. Frost, K. T. Butler, F. Brivio, C. H. Hendon, M. van Schilfgaarde, A. Walsh, *Nano Letters* **2014**, 14, 2584.
[60] H.-W. Chen, N. Sakai, M. Ikegami, T. Miyasaka, *The Journal of Physical Chemistry Letters* **2015**, 6, 164.
[61] J. Heyd, J. E. Peralta, G. E. Scuseria, R. L. Martin, *The Journal of Chemical Physics* **2005**, 123, 174101.
[62] J. Heyd, G. E. Scuseria, *The Journal of Chemical Physics* **2004**, 121, 1187.
[63] H. Xiao, J. Tahir-Kheli, W. A. Goddard, *The Journal of Physical Chemistry Letters* **2011**, 2, 212.
[64] M. Henderson Thomas, J. Paier, E. Scuseria Gustavo, *physica status solidi (b)* **2010**, 248, 767.
[65] S. P. Ong, W. D. Richards, A. Jain, G. Hautier, M. Kocher, S. Cholia, D. Gunter, V. L. Chevrier, K. A. Persson, G. Ceder, *Comp Mater Sci* **2013**, 68, 314.
[66] *NIST Standard Reference Database No. 69, edited by P.J. Linstrom and W.G. Mallard (National Institute of Standards and Technology, Gaithersburg, MD, 2003).*
[67] R. M. Jacobs, J. H. Booske, D. Morgan, *Phys Rev B* **2012**, 86, 054106.
[68] R. Jacobs, B. Zheng, B. Puchala, P. M. Voyles, A. B. Yankovich, D. Morgan, *The Journal of Physical Chemistry Letters* **2016**, 7, 4483.
[69] *Reference Solar Spectral Irradiance: ASTM G-173:* http://rredc.nrel.gov/solar/spectra/am1.5/astmg173/astmg173.html **2017**.